\documentclass[journal=jacsat,manuscript=article]{achemso}

\usepackage[version=3]{mhchem} 
\usepackage{xcolor}
\usepackage{amsmath}
\usepackage{siunitx}
\usepackage{color,soul}
\usepackage{overpic}
\let\vec\mathbf

\author{Khatereh Azizi}
\affiliation{The Abdus Salam International Centre for Theoretical Physics, Strada Costiera 11, 34151 Trieste, Italy}
\alsoaffiliation{Quantum Biology Laboratory\footnote{\url{https://quantumbiolab.com}}, Howard University, Washington, DC 20060, USA}
\author{Matteo Gori}
\affiliation{Quantum Biology Laboratory\footnote{\url{https://quantumbiolab.com}}, Howard University, Washington, DC 20060, USA}
\author{Uriel Morzan}
\affiliation{The Abdus Salam International Centre for Theoretical Physics, Strada Costiera 11, 34151 Trieste, Italy}
\author{Ali Hassanali}
\affiliation{The Abdus Salam International Centre for Theoretical Physics, Strada Costiera 11, 34151 Trieste, Italy}
\author{Philip Kurian}
\affiliation{Quantum Biology Laboratory\footnote{\url{https://quantumbiolab.com}}, Howard University, Washington, DC 20060, USA}
\email{pkurian@howard.edu}

\title[]
  {Examining the origins of observed terahertz modes from an optically pumped atomistic model protein in aqueous solution
  }

\keywords{American Chemical Society, \LaTeX} 

\begin{document}







\begin{abstract}

The microscopic origins of terahertz (THz) vibrational modes in biological systems are an active and open area of current research. Recent experiments [\textit{Physical Review X} \textbf{8,} 031061 (2018)] have revealed the presence of a pronounced mode at $\sim$0.3 THz in fluorophore-decorated bovine serum albumin (BSA) protein in aqueous solution under nonequilibrium conditions induced by optical pumping. This result was heuristically interpreted as a collective elastic fluctuation originating from the activation of a low-frequency phonon mode. In this work, we show that the sub-THz spectroscopic response emerges in a statistically significant manner ($>2\sigma$) from such collective behavior, illustrating how specific THz vibrational modes can be triggered through optical excitations and other charge reorganization processes. We revisit the theoretical analysis with proof-of-concept molecular dynamics that introduce optical excitations into the simulations. Using information theory techniques, we show that these excitations can induce a multiscale response involving the two optically excited chromophores (tryptophans), other amino acids in the protein, ions, and water. Our results motivate new experiments and fully nonequilibrium simulations to probe these phenomena, as well as the refinement of atomistic models of Fr\"{o}hlich condensates that are fundamentally determined by nonlinear interactions in biology.

\end{abstract}

\section{Significance Statement}
Recent spectroscopic experiments have suggested that photoexcited biomolecules can behave like optomechanical transducers, with potentially giant dipole moments oscillating in the sub-THz domain. A heuristic explanation of these observations has been that the photoexcitation activates a collective vibration through an energy downconversion cascade in the biomolecule, resulting in energy ``funnelling'' to a specific elastic mode. However, a fuller understanding of this activation mechanism is still lacking. The work we present explores how the mechanical properties of a solvated model protein at thermodynamic equilibrium are affected by the photoexcitation of intrinsic chromophores through molecular dynamics simulations. In particular, we show how optical excitations and consequent charge reorganization processes can trigger sub-THz collective vibrational modes. Moreover, we show how photoexcitation can induce a multiscale response involving not only the amino acids in the protein but also the surrounding counterions and water.

\section{Introduction}

In the last decade, there has been an increased interest in studying the picosecond (terahertz, THz) regime of biological systems\cite{markelz2011}. Both proteins and the aqueous solvent have a distinct absorption in the THz regime\cite{havenith2014,heyden2010,falconer2012,falconer2014,donnan2015}, extending from the sub-THz to several tens of THz. However, advances in THz Fourier-transform infrared (FTIR), time-domain (TDS), and continuous-wave (CW) near-field spectroscopies have illuminated the variety of vibrational modes and timescales associated with the protein-water interface \cite{hassanali2016,havenith2007,havenith2008,havenith2009,alfano2004}, which has revealed correlated dynamics occurring at much larger distances than previously thought possible in a range of proteins \cite{donnan2015}. However, the molecular origins of these vibrational modes still remain elusive. Some reports have suggested the possible relevance of these modes to fundamental biological behaviors far from thermal equilibrium, including in bacterial cytoplasm \cite{jacobs-wagner2014}, brain tissue \cite{alfano2016}, and conscious processing \cite{penrose1994,hush2009,kurian2017,kurian2018,kurian2019}.


\begin{figure}[h!]
    \centering
    \includegraphics[scale=0.32]{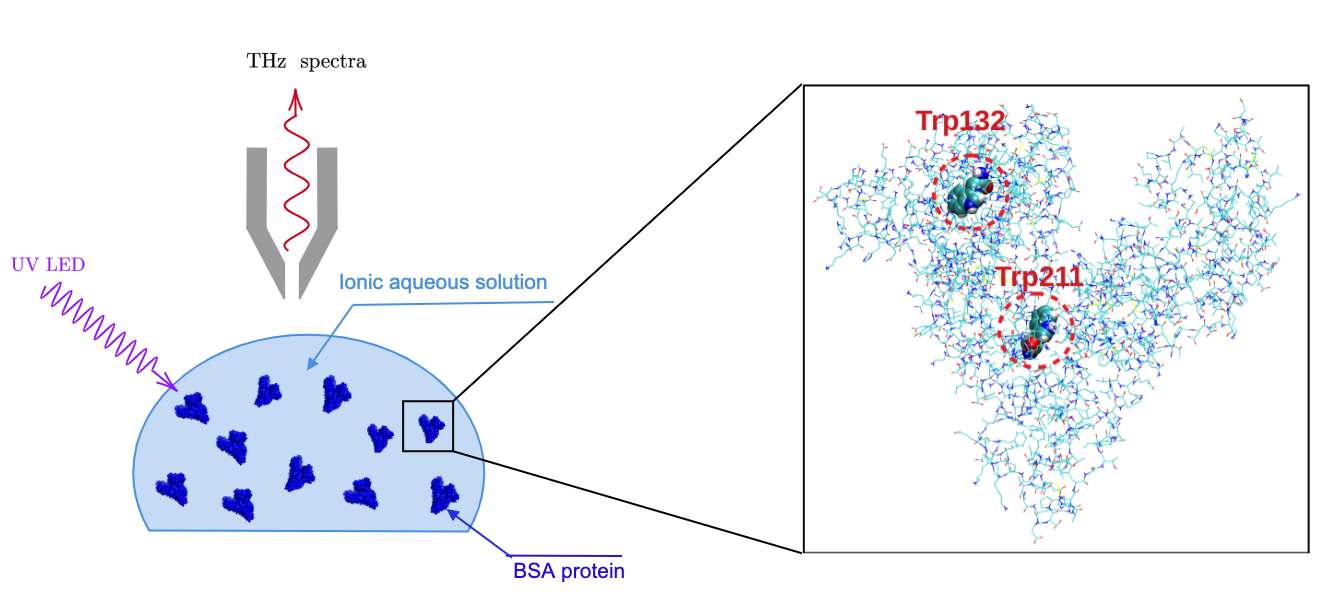}\\
    \includegraphics[scale=0.35]{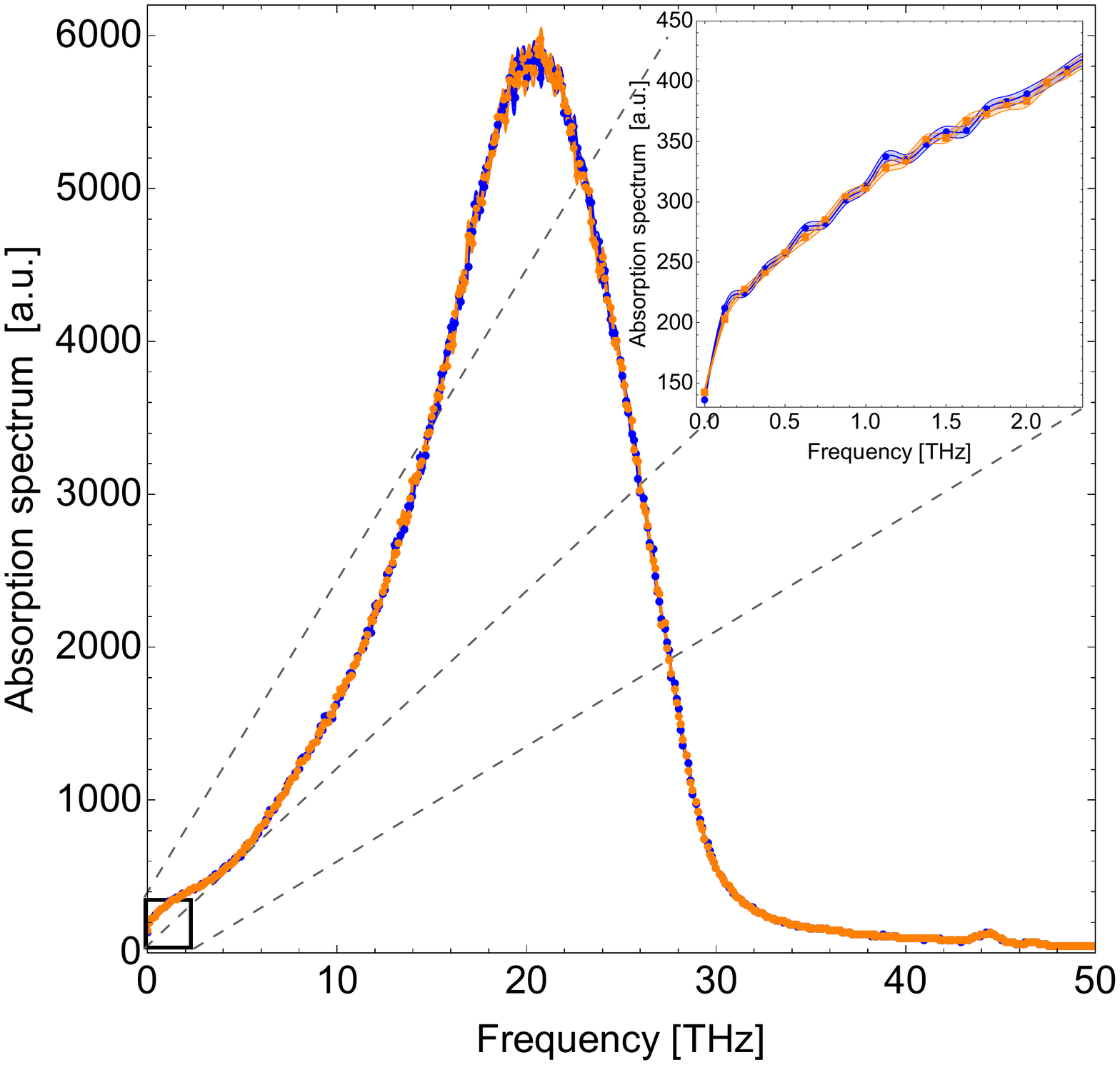}\\
    \caption{Left upper panel: A schematic presentation of the experiment \cite{nardecchia2018out} studying the terahertz spectra of an aqueous protein solution, which we revisit by atomistic simulations in our study. Right upper panel (zoomed-in picture): Atomic structure of the protein bovine serum albumin (BSA). Tryptophan residues, the dominant ultraviolet light-absorbing amino acid in proteins, are critical for the excited-state effects observed in the entire BSA and are highlighted here. Water and ions are omitted for clarity.
    Bottom panel: Simulated absorption coefficient $\alpha_{\rm GS(EX)}(\nu)$ (in arbitrary units, a.u.) of the whole system vs. frequency $\nu$ for the BSA aqueous solution, with tryptophans in the ground state (blue) and the photoexcited state (orange). In the inset, the detail of the absorption spectrum of the protein-ions-water system is shown, highlighting the low-frequency range $(0-2.35 \rm{THz})$.} 
    \label{fig:THzspec}
\end{figure}

In a recent work bridging theory, computation, and experiment\cite{nardecchia2018out}, the model protein bovine serum albumin (BSA) was decorated with Alexa488 fluorophores and pumped far from equilibrium in aqueous solution (see Figure 1), thereby mimicking biological conditions, by energy injection from both a blue 488-nm laser and a broadband ultraviolet light-emitting diode (UV LED) \cite{MPettiniprivcomm}. While the blue laser certainly served to excite the Alexa488 fluorophores (which are known to bind covalently to lysine residues), the UV LED centered at 255 nm excited the two tryptophan residues in BSA, which have their peak absorption at 280 nm and a strong transition dipole moment in this spectral range. Two experimental groups independently found an absorption peak at $\sim\SI{0.314}{\,\tera\hertz}$ under these nonequilibrium (NEQ) conditions. (The solution was thermostatted, and therefore the NEQ state of each protein molecule is a stationary one, loosely speaking, as in the case of a lasing medium where pumping and dissipation coexist.) This absorption peak vanished if the fluorophores were removed or if the external pumping ceased. The feature could also be recovered by applying the energy injection again for a sufficiently long time, owing to the dissipative losses of the medium. Heuristic calculations \cite{bastrukov1994} modelling BSA as an elastic sphere predict the lowest-frequency breathing mode
at $\SI{0.308}{\,\tera\hertz}$ with a precision of $\sim 2\%$ with respect to the observed absorption peak. This line of thought and evidence have been interpreted in terms of the emergence of a pronounced collective phonon mode in the aqueous BSA system due to the activation of a Fr\"{o}hlich-like condensation. Such unexpected protein behavior has been observed in a warm and wet environment, demanding a deeper study of its origins and the sub-THz response in these conditions, which remain poorly understood. 

The possibility of activating giant coherent dipole oscillations in biomolecules is a long-standing proposal, one which may be realized by the channelling of a large part of the vibrational energy of a biomolecule into its lowest-frequency normal modes. Two conditions are required to manifest such a phenomenon: a sufficiently high energy injection rate, and nonlinear interactions among the phononic normal modes mediated by a thermal bath. These phenomena have been studied in quantum\cite{frohlich1968long,wu1978bose,pokorny1998biophysical,pokorny1992heat,turcu1997generic,zhang2019quantum}, semi-classical\cite{preto2017semi}, and classical\cite{nardecchia2018out} frameworks, and such giant dipole oscillations in the sub- to few-THz range could have a major impact on the description of biomolecular dynamics. It has already been theoretically proposed \cite{preto2015possible} and very recently experimentally realized \cite{lechelon_experimental_2022} that resonance among these oscillating dipoles is responsible for (selective) long-range electrodynamic interactions, which could play a relevant role in recruiting cognate partners involved in a range of biochemical reactions. 


First experimental evidences\cite{lundholm2015terahertz} of Fr\"{o}hlich-like condensation were observed in lysozyme crystals with X-ray crystallography under an energy injection provided by $0.4$-THz illumination. In these experiments, the electron density along the $\alpha$-helices revealed that the thermalization of low-frequency 
vibrations occurs on timescales of micro- to milliseconds, much longer than the expected nanosecond time scale. This picture is consistent with a Fr\"{o}hlich-like condensation of the energy into the lower-frequency phononic modes. In a recent letter \cite{elsaesser2021}, strong Coulombic interactions between free electrons and their solvating water shells resulted in commensurate electronic-nuclear coupling and a polaronic response characterized by the excitation of low-frequency vibrational and librational modes of the water environment. In that work, the time-resolved change of the dielectric function mapped by broadband THz pulses demonstrated pronounced oscillations persisting up to 30 ps, with their frequency increasing with electron concentration from 0.2 to 1.5 THz. These intimate relationships between electronic states dressed by longitudinal excitations of the environment have been suggested for decades \cite{vitiello1985,vitiello1986,vitiello1988} and applied in recent years to understanding DNA-protein interactions \cite{kurian2018}.

Spectroscopically active low-frequency (0-3 THz) vibrational modes of proteins have been implicated indirectly in enzyme catalysis and other functions \cite{cheatum2020}, suggesting an intimate relationship between charge reorganization processes and collective protein vibrations. For example, a computational study of human heart lactate dehydrogenase \cite{schwartz2011} identified a preferred vibrational mode through which thermal energy can be channeled due to protein anisotropies in the active site. Soybean lipoxygenase, a prototypical enzyme system where thermal activation of the protein scaffold by solvent collisions has been linked to the efficiency of active site chemistry \cite{klinman2019}, raises intriguing possibilities regarding the role of downstream signaling via thermally or light-induced protein conformational change. Such activation of particular ``hot'' residues, and charting the potentially rate-promoting dissipation channels through protein systems, is a lively area of current research \cite{faraji2021}.

The experimental setup shown in the upper panels of Figure \ref{fig:THzspec} leading to a large peak at around 0.3 THz requires UV excitation. It is well known that one of the most important intrinsic chromophores in proteins is trytophan, which absorbs broadly in the UV with a peak at around 280 nm and fluoresces generally near 350 nm in aqueous solution \cite{chergui2010}. Tryptophan excited states can serve as excellent reporters for the local electrodynamic behaviors of biological systems \cite{callis1997,chergui2009}. Numerous experiments, simulations, and theory have shown that the photoexcitation of trytophan in water as well as in biomolecules involves a complex response of water, ions, and also the surrounding protein on timescales ranging from several to hundreds of picoseconds\cite{Qin8424,Qiu13979,singer2007,halle2009}. 

The BSA protein that we study in this paper has two tryptophans, both of which can be excited by UV light. In the previous study by Pettini and coworkers, the role of the tryptophans as optically excitable centers was enhanced by the addition of Alexa488 fluorophores, which absorb mainly blue light. Since tryptophans are excited by more energetic UV excitations, they may play a critical role in stimulating the observed THz modes after energy downconversion through the protein. Furthermore, from the point of view of taking a first step at using hybrid classical atomistic models with excited-state charge configurations to understand the origins of Fr\"{o}hlich-like condensates, modeling the naturally occurring tryptophan chromophores is more computationally tractable and more widely applicable to other protein systems. The interpretation of these experiments within the framework of a Fr\"{o}hlich-like condensate considers the tryptophans as essential to optical pumping, initiating the downconversion process from photon excitation to phonon vibration through the protein, but such a framework provides only a coarse-grained perspective on the water, ions, and protein response. Tryptophan plays dual roles in this phenomenon, as (1) highly efficient absorbers of energetic UV photons, which can change whole-system response properties in the electronic degrees of freedom at thermal equilibrium; and as (2) centers of energy downconversion through nonlinear interactions with the protein and aqueous environment, which when pumped far from thermal equilibrium can populate protein vibrational modes in such a manner that violates energy equipartition. In this article we consider only the first of these roles, which is an essential prerequisite to analyzing the second from an atomistic point of view. 

It should be noted that all aromatic residues, and certain other complexes in proteins, absorb in the UV range, which can affect biomolecular photophysics and resulting downstream responses. Tyrosine and phenylalanine, which generally have smaller transition dipole moments than tryptophan in the UV, frequently show up in proteins, including BSA, as a larger proportion of the total number of amino acids. There is some additional evidence that a visible fluorophore in close proximity to a lysine residue (namely, the experimental scenario with Alexa488 attached to BSA, which inspired the present work) absorbs significantly in the UV upon Schiff-base deprotonation \cite{rousseau1996,salcedo2003}. 

A refined description of the dynamics that allows for the emergence of NEQ phononic terahertz modes far from energy equipartition---a key feature of Fr\"{o}hlich condensation---is still lacking.
Moreover, to derive the Fr\"{o}hlich rate equations from simple idealized models of quantum (or classical) harmonic oscillators, many assumptions have
 been considered that often do not hold in real experiments. For instance, it is usually assumed that the characteristic timescale of the dynamics of the occupation number of the collective modes of the protein is much larger than the inverse of their frequencies, so that they are completely decoupled. However, it has been found that the experimentally measured quality ($Q$) factor of the absorption peak associated with the NEQ giant dipole oscillations is $\sim$50-90\cite{gori:hal-03257350}. This means that the coherence time of such activated collective modes is estimated to be $\tau_{\text coherence}=Q/(\pi\nu)\approx 16-30$ ps\cite{lechelon_experimental_2022}. Comparing the full width at half-maximum of the absorption peak observed in photoexcited BSA to the unexcited one, we can estimate $Q \approx 30$. This means that the characteristic timescale of the observed coherent oscillations is very close to the timescale of the characteristic BSA breathing mode that is conjectured to be responsible for the observed signal. The actual limitations of existing analytical and numerical models of Fr\"{o}hlich-like condensation have led us to investigate the details of such phenomena in realistic biomolecular systems through atomistic molecular dynamics simulations.

In this work, we revisit the analysis of the optically pumped BSA experiments using classical atomistic molecular dynamics in thermal equilibrium, taking first steps in providing molecular-level information that can be used to refine existing theory on the microscopic origins of the phenomenon. Indeed, at the heart of the Fr\"{o}hlich rate equations is the nature of the crucial nonlinear couplings with the external pumping field and with the thermal bath, which are not generally modeled with classical molecular dynamics. Nevertheless, as an essential milestone en route to NEQ simulations with such nonlinear interactions included, insights can be gleaned from the standard approach. Herein, we use a simplified, computationally tractable atomistic model to investigate the mechanical perturbations that can occur when electronically excited tryptophan residues in the BSA protein are simulated with classical molecular dynamics simulations in thermal equilibrium. We find that UV photoexcitation of tryptophan introduces large structural and dynamical changes in BSA involving regions far away from the two excited residues, suggesting the importance of a potential nonlinear response in the system. Furthermore, we also observe preliminary signatures of nonlinear response in the water and surrounding ions. By computing the terahertz spectra from the ground- and photoexcited-state BSA simulations, we illustrate the possibility of exciting specific THz modes emerging from a collective multiscale response of the protein, ions, and water.

\section{Computational Methods}

\subsection*{Classical Molecular Dynamics}

All molecular dynamics simulations were performed using the GROMACS code\cite{abraham2015}. The initial crystal structure was obtained from the Protein Data Bank (PDB, with ID 3V03)\cite{majorek2012structural}. Following energy minimization, we subsequently performed $NVT$ (isothermal-isochoric) and $NPT$ (isothermal-isobaric) equilibrations each for 500 ps. For these runs, the temperature was set to 300K and a pressure of 1 bar was used in the $NPT$ ensembles. After these steps, the simulations were run within the $NVT$ ensemble for approximately 0.5 microsecond in either the ground or optically excited state, as will be explained in detail later.

The GROMOS96 54a7\cite{Schmid2011} force field was used for the simulation together with the SPC/E\cite{spcemodel} water model. Bond lengths were constrained using LINCS\cite{lincs1997}.
The protonation states of the amino acids were chosen to be those relevant to neutral pH, which leads to a net negative charge of $-16e$ in the protein. The system is thus neutralized by adding 16 sodium ions. Additional sodium and chloride ions were added, leading to a buffer solution of concentration 0.15~M. Long-range electrostatics were calculated using the particle mesh Ewald (PME) method\cite{pme}. The cutoff for both pairwise van der Waals (vdW) and short-range electrostatic interactions was set at 1.2 nm\cite{piana_evaluating_2012}. Analytical corrections for pairwise interactions beyond the vdW cutoff, as implemented in GROMACS, were used to correct for the total energy and pressure of the system. The simulations were done in a cubic box of side length of $\sim$13.3 nm, with periodic boundary conditions, using a timestep of two femtoseconds. The total number of water molecules in the system was 75,414.

\subsection*{Photoexcitation Model}
\color{black}
Simulations of BSA were performed both in the ground (Gr) and excited (Ex) state. The excited state of BSA was modeled by electronically exciting its two tryptophan (Trp) residues (with residue numbers 132 and 211). The two lowest-lying transition dipole states for tryptophan are usually referred to as the $\text{L}_\text{a}$ and $\text{L}_\text{b}$ states, which demonstrate polarization along the short and long (orthogonal) axes in the $\pi-\pi^*$ orbital, respectively (see Figure S1). $\text{L}_\text{a}$ is generally the lower-lying fluorescent state in Trp in solvated proteins, and in water there is ultrafast internal conversion (on the order of tens of femtoseconds) from $\text{L}_\text{b}$ to $\text{L}_\text{a}$ upon absorption, which is not modeled. We focus instead on the BSA protein relaxation and response due to both tryptophans being in the excited $\text{L}_\text{a}$ state.

The ground-state charges are those of the GROMOS96 54a7 forcefield\cite{Schmid2011}. The photoexcitation is modeled by modifying the point charges of the indole-group atoms in both Trp residues in the protein. The charge differences between ground and excited state are based on multi-reference quantum chemistry (CASSCF/CASPT2) electronic structure calculations performed by Sobolewski and coworkers \cite{sobolewski1999ab}. This procedure has been applied in several previous studies in our group and has been found to give good agreement with experiments\cite{hassanalikwk2006,singer2007}. 

\subsection*{Terahertz Spectra Analysis}

To compute the terahertz spectra we employed a time correlation formalism based on dipole fluctuations sampled in molecular dynamics simulations \cite{mcquarrie_statmech_bk, bornhauser2001intensities}. This entails extracting the Fourier transform of the dipole-dipole time correlation function.
The total dipole moment of the system is given by contributions coming from the protein ($\vec{M}_P(t)$), water ($\vec{M}_W(t)$), and ions ($\vec{M}_I(t)$). Specifically, the absorption coefficient per unit length is given by 

\begin{equation}
\label{eq:abs_coeff}
    \alpha(\nu)= F(\nu)\int_{-\infty}^{\infty} dt \, e^{-i2\pi \nu t} \big \langle \frac{d}{dt}\vec{M}(0) \cdot
    \frac{d}{dt}\vec{M}(t)
    \big \rangle,
\end{equation}

\begin{equation}
   \vec{M} = \vec{M}_P + \vec{M}_W
   + \vec{M}_I,
\end{equation}
where $F(\nu) = (1/4\pi\epsilon_0)(2\pi/(k_BT))(1/(3Vcn(\nu)))$, with frequency $\nu$, refractive index $n(\nu)$, volume $V$, and temperature $T$.

Note that the terahertz spectra includes contributions coming from both auto-correlations and cross-correlations involving the protein, water, and ions. This will be discussed in more detail later in the manuscript. In order to compute the terahertz spectra, a total of $\sim$200 configurations were selected from both the long ground- and excited-state runs described earlier. From each of these initial configurations, microcanonical dynamics (holding particle number, volume, and energy constant) were performed for 100 ps. The terahertz spectra were then averaged over all these trajectories. This protocol has been previously applied with success to both protein and water systems\cite{heyden2013spatial}.

Experimental evidence\cite{nardecchia2018out} of collective phononic modes in biosystems driven far from
thermal equilibrium is observed in sub-terahertz difference spectra between the
photoexcited and ground states of BSA protein in aqueous solution. To ascertain the statistical relevance of the 
observed signals in molecular dynamics simulations,
we assume the difference in the absorption coefficient of Equation \eqref{eq:abs_coeff} between the photoexcited state and the ground state follows a Gaussian distribution at each frequency. 

In order to identify differences in the various terahertz spectral contributions from different modes upon photoexcitation, we 
quantify statistically significant changes as those exceeding two times the standard deviation,

\begin{equation}
    \sigma(\nu)=\sqrt{\sigma_{\alpha_{\text{Ex}}}^2(\nu)+\sigma_{\alpha_{\text{Gr}}}^2(\nu)},
    \label{eq:standdev}
 \end{equation}%
where $\sigma_{\alpha_{\text{Ex(Gr)}}}(\nu)$ is the statistical error over the absorption
coefficients $\alpha$ in the photoexcited (ground) state. 
Signals in $\Delta \alpha (\nu)=\alpha_{\text{Ex}}(\nu)-\alpha_{\text {Gr}}(\nu)$ have been considered statistically significant if $|\Delta \alpha (\nu)|$ exceeds $2\sigma(\nu)$, such that random fluctuations would not account for these outlier values in at least 95\% of cases.

\subsection*{Centrality Analysis}

In order to characterize the dynamical changes induced by photoexcitation, we performed a mutual information (MI)-based eigenvector centrality and degree centrality analysis. These centrality metrics are widely employed to capture fundamental properties of networks. Negre et al. have recently demonstrated the capability of the eigenvector centrality measure based on the mutual information to disentangle the complex interplay of
amino acid interactions giving rise to allosteric signaling \cite{NegreE12201}. 
This measure enables pinpointing of the key residues involved in the main collective motions of the system. 
To perform the centrality analysis we map the protein MD trajectory onto a graph composed of nodes interconnected by edges. Each node represents a single atom in the system,  
and the edges connecting pairs of these nodes are defined as the dynamical cross-correlations experienced by the corresponding atoms along an MD trajectory. The latter can be quantified employing the generalized correlation coefficients, based on the MI between two atoms ${\text{G}}_{ij}$:
\begin{equation}
   \label{G}
    {\text{G}}_{ij}= \Bigg[1-\text{exp}\Bigg(-\frac{2}{3}\,\text{I}[{\bf x}_i , {\bf x}_j ]\Bigg)\Bigg]^{\frac{1}{2}},
\end{equation}
where ${\bf x}_i$ are the zero-mean atomic displacement vectors extracted from the MD simulations. The MI between the two atoms, $\text{I}[{\bf x}_i , {\bf x}_j ]$, is defined in the following way:
\begin{equation}
    {\text{I}}[{\bf x}_i , {\bf x}_j ]= H_i+H_j - H_{ij},
\end{equation}
where 
\begin{equation}
    H_k = \int  \text{p}({\bf x}_k) \text{ln}\,\text{p}({\bf x}_k)d{\bf x}_k
\end{equation}
\begin{equation}
    H_{ij} = \int \int  \text{p}({\bf x}_i,{\bf x}_j) \text{ln}\,\text{p}({\bf x}_i,{\bf x}_j)d{\bf x}_i d{\bf x}_j
\end{equation}
are the marginal and joint Shannon entropies, respectively, extracted from MD simulations of the system at equilibrium. The generalized correlation coefficient
${\text{G}}_{ij}$ ranges from zero to one for uncorrelated to fully correlated variables, respectively. We have recently shown that the compression of the generalized correlation matrix into a centrality vector enables the subtle characterization of the role played by every atom in the major collective modes of the system. 

The eigenvector centrality metric E$_c$ was obtained employing the linear approximation of the mutual information introduced by Lange et al.\cite{Lange2006} 
The linear mutual information (LMI) relies on a Gaussian approximation of the probability density, which can be expressed as

\begin{equation}
    p({\bf x}_i,{\bf x}_j) = \frac{1}{(2\pi)^3\text{det} ( \text{Cov})} \exp \Bigg[ -\frac{1}{2} ({\bf x}_i,{\bf x}_j)\text{Cov}^{-1} ({\bf x}_i,{\bf x}_j)^T   \Bigg],
\end{equation}
where $\text{Cov} = \langle ({\bf x}_i,{\bf x}_j )^T \otimes ({\bf x}_i,{\bf x}_j ) \rangle$ is the covariance matrix, and ${\bf x}_i, \, \, {\bf x}_j,\text{ and }( {\bf x}_i, {\bf x}_j)$ are row vectors, with $({\bf x}_i, {\bf x}_j)$ defined as a concatenation of vectors. Marginal probabilities are computed from the marginal covariance matrices $\text{Cov}_{i}= \langle {\bf x}_i^T \otimes {\bf x}_i \rangle $.  Hence the entropies $H$ are obtained analytically from the Gaussian density approximations as
\begin{equation}
    H =\frac{1}{2} [q+\text{ln} (\text{det}(\text{Cov}))],
\end{equation}
where $q=2d(1+ \text{ln} (2\pi) )=17.03$, and $d=3$ is the dimension of each individual coordinate vector. The LMI is then defined as 
\begin{equation}
\text{LMI}_{ij} = \frac{1}{2} \Big[ \Big(\text{ln}( \text{det}(\text{Cov}_i ))\Big)  +\Big(\text{ln}( \text{det}(\text{Cov}_j ))\Big) -\Big(\text{ln}(\text{det}(\text{Cov} ))\Big)  \Big].  
\end{equation}

As discussed above, the MI-based E$_c$ distribution is then obtained from the leading eigenvector of the G matrix (Equation \ref{G}, where $\text{I}[{\bf x}_i, {\bf x}_j ]$ is approximated by $\text{LMI}_{ij}$), which by virtue of the Perron-Frobenius theorem is unique (i.e., non-degenerate):

\begin{equation}
    E_c(i)=\frac{1}{\lambda}\sum_j \text{G}_{ij} \text{E}_c(j).
\label{eq:11}
\end{equation}

We would like to highlight that the covariant matrices $\mathrm{Cov} $ and $\mathrm{Cov}_i$ are based on time (and statistical) averages of
correlations among the displacements $\mathbf{x}_{i}$.
From this follows that the correlation matrices are based on the zero-frequency ($\omega=0$) correlations in the frequency domain. 
It is possible, in principle, to construct a centrality score based on frequency-filtered correlations among atomic displacements $\text{Cov}(\omega) = \langle ({\bf x}_i,{\bf x}_j )^T \otimes ({\bf x}_i,{\bf x}_j ) e^{i\omega t}\rangle$ and $\text{Cov}_i(\omega) = \langle {\bf x}_i^T \otimes {\bf x}_i e^{i\omega t}\rangle$. Such an analysis is beyond the scope of the present paper and will be developed in further works. 






\section{Results}

Here we elucidate the mechanisms by which the photoexcitation of the BSA protein can induce a collective response involving various modes in the system. In particular, we will begin by identifying specific changes in the sub-terahertz regime involving the protein, water, and ions. We will then subsequently illustrate structural and dynamical changes that involve these various contributions.


\color{black}

\subsection{Origins of sub- and few-terahertz modes}

The bottom panel of Figure \ref{fig:THzspec} shows the total absorption spectra of the system consisting of all the auto- and cross-correlations of the protein, ions, and water in the range between $\sim 0-30$ THz. Since the specific focus of our work is to examine more deeply the region between $\sim 0-1$ THz, we show a zoomed inset. Indeed, there is a significant contribution of intensity in this region as well.

As alluded to earlier, the terahertz signal can arise from a complex coupling of various modes in the system. There have been numerous theoretical studies in the literature illustrating how low-frequency terahertz spectra (see inset of Figure \ref{fig:THzspec} bottom panel) can arise from the interplay of protein, ion, and water fluctuations. Here we find that the photoexcitation of the two tryptophan chromophores in BSA gives rise to a multiscale response involving various different contributions.

\begin{figure}[ht!]
    \centering
    \includegraphics[scale=0.35]{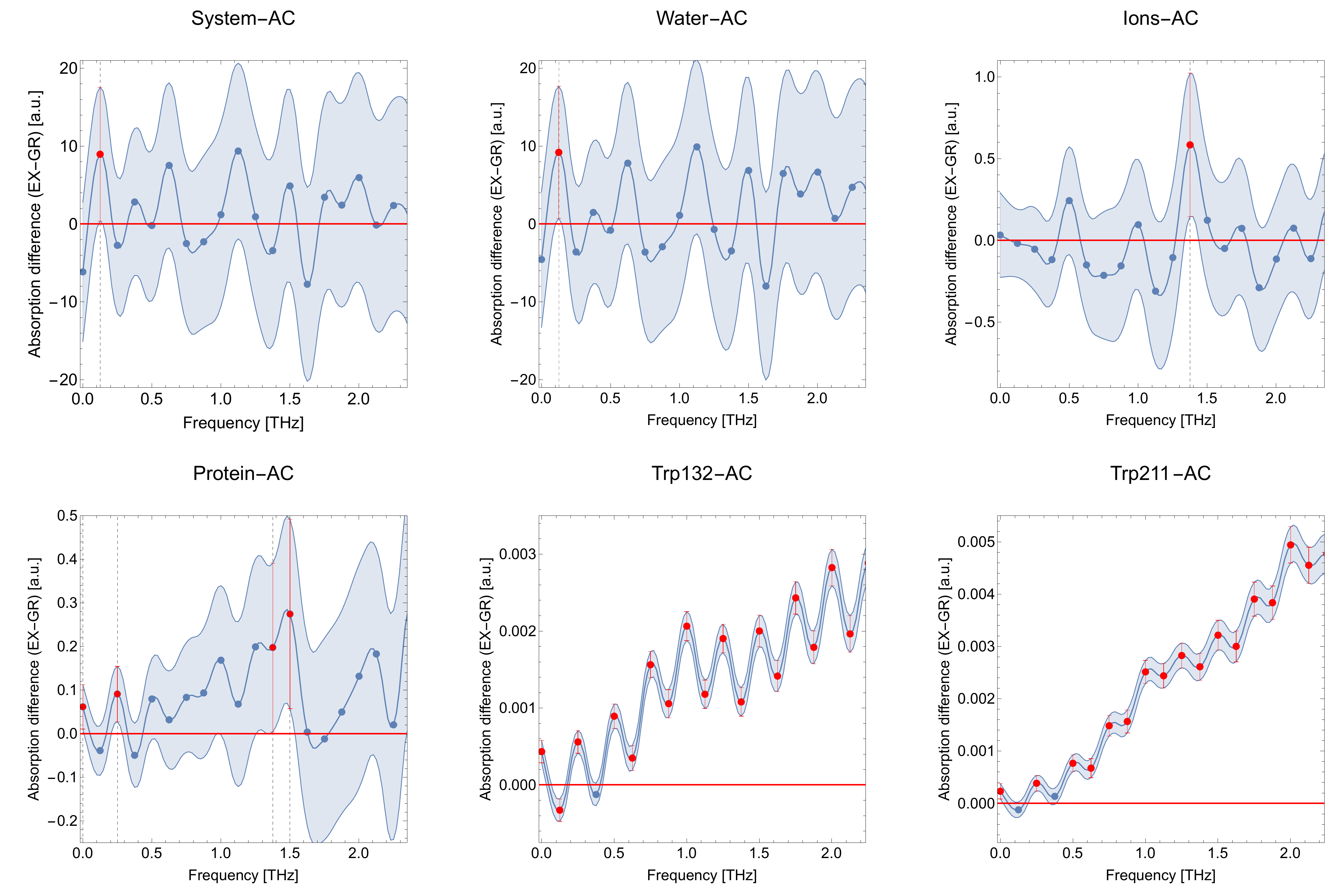}
    \caption{Autocorrelations (AC) of electric dipole fluctuations from different inclusions of the system (top row: whole system, water, ions; bottom row: protein, tryptophan 132, tryptophan 211) contributing to the simulated
    terahertz absorption spectra in the range $0-2.35\,\,\rm{THz}$ (Figure \ref{fig:THzspec}) upon
    photoexcitation of bovine serum albumin (BSA) protein, presented as differences in the absorption between photoexcited (Ex) and ground (Gr) state simulations ($\Delta\alpha(\nu) = \alpha_{\text{Ex}}(\nu) - \alpha_{\text{Gr}}(\nu)$).
    The shaded band above and below the blue data points is an interpolated second-order polynomial and has a width at each data point 
    of four times the standard deviation,
    $\Delta\alpha(\nu)\pm 2\sigma (\nu)$. The black vertical dashed lines extending from the bold red error bars correspond to the frequencies where the
    absorption difference exceeds $2\sigma$-statistical significance
    (i.e., when the mean value $\Delta \alpha$ is separated from zero (x-axis, in bold red) by more than $2\sigma$).
    Vertical dashed lines have not been reported for the two tryptophan AC spectra for the sake of readability. The simulated terahertz absorption spectra from which these differences were derived have each been obtained by averaging over 211 $NVE$ trajectories holding particle number, volume, and energy constant.}
    \label{fig:thz1}
\end{figure}

\begin{figure}[ht!]
    \centering
    \includegraphics[scale=0.35]{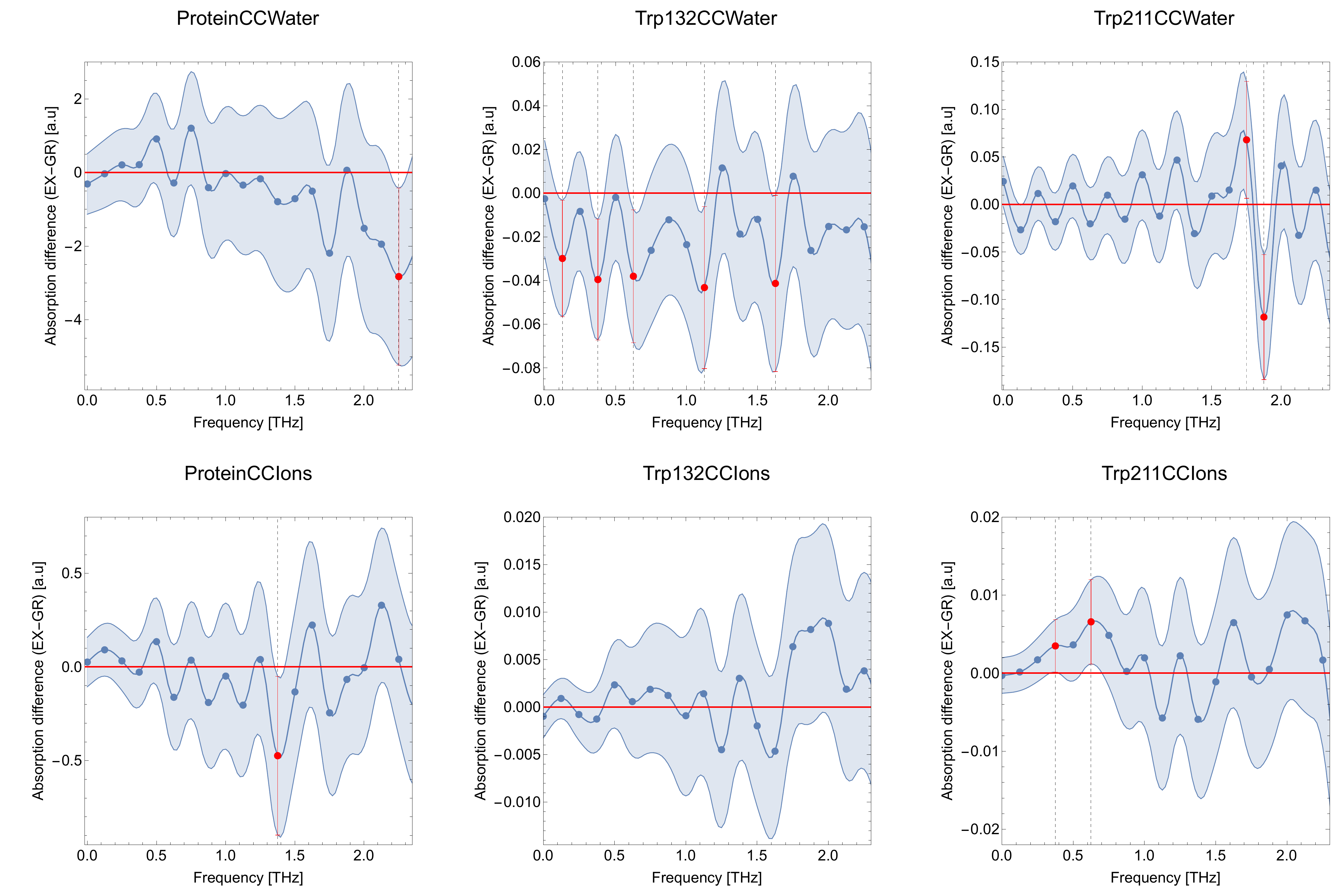}
    \caption{Cross-correlations (CC) of electric dipole fluctuations from different subpartition partners of the system (top row: protein-water, tryptophan 132-water, tryptophan 211-water; bottom row: protein-ions, tryptophan 132-ions, tryptophan 211-ions) contributing to the simulated
    terahertz absorption spectra in the range $0-2.35\,\,\rm{THz}$ (Figure \ref{fig:THzspec}) upon
    photoexcitation of bovine serum albumin (BSA) protein, presented as differences in the absorption between photoexcited (Ex) and ground (Gr) state simulations ($\Delta\alpha(\nu) = \alpha_{\text{Ex}}(\nu) - \alpha_{\text{Gr}}(\nu)$).
    The shaded band above and below the blue data points is an interpolated second-order polynomial and has a width at each data point 
    of four times the standard deviation,
    $\Delta\alpha(\nu)\pm 2\sigma(\nu)$. The black vertical dashed lines extending from the bold red error bars correspond to the frequencies where the
    absorption difference exceeds $2\sigma$-statistical significance
    (i.e., when the mean value $\Delta \alpha$ is separated from zero (x-axis, in bold red) by more than $2\sigma$).
    Vertical dashed lines have not been reported for the tryptophan 132-ions CC spectra because there are no statistically significant points, reflecting the greater sensitivity of tryptophan 211 to photoexcitation-induced conformational changes in the ionic environment (see Figure S3). The simulated terahertz absorption spectra from which these differences were derived have each been obtained by averaging over 211 $NVE$ trajectories holding particle number, volume, and energy constant.}
    \label{fig:thz2}
\end{figure}

In equilibrium measurements of THz spectra in the low-frequency regime, it is common to observe broad linewidth absorptions, rather than sharp peaks at specific frequencies. However, the BSA experiments reporting an absorption peak at $\sim 0.3$ THz are conducted in nonequilibrium settings. Since we cannot make a direct comparison at the moment in our simulations, we focus in the next section on reporting the differences in the bare intensities of the spectra between ground and excited states, without applying any smoothening procedures to the spectra. Significant peaks are seen at points throughout the low-frequency range, which contribute to the observed elastic mode of the whole BSA protein in ionic solution. This elastic mode is constrained by the intrinsic vibrational properties (Young's modulus, hydrodynamic radius, specific volume) of the protein. \cite{nardecchia2018out}

Figure \ref{fig:thz1} shows the differences in the autocorrelation contributions to the terahertz absorption differences between excited and ground states, from various components of the protein solvated with ions and water as described earlier in the manuscript. In particular, the top row of panels in Figure \ref{fig:thz1} displays the differences in the autocorrelations arising from the whole system, from water, and from ions. Though compared to the Trp132 and Trp211 autocorrelations there are far fewer data points across the entire low-frequency THz range that are significantly different from zero ($|\Delta \alpha(\nu) |>2\sigma (\nu)$), we observe that for the whole system, for the water, for the ions, and for the protein, the region between 0 and 2.35 THz is characterized by some rather large autocorrelation changes upon photoexcitation (significant at the $2\sigma$-level, twice the usually reported error, at 95\% or greater confidence interval). Particularly notable are the statistically significant autocorrelation changes in the sub-THz region for the protein, for the water, and for the whole system, suggesting pathways for sub-THz response of the whole system that has been observed experimentally under NEQ conditions.

We observe in the bottom row of panels in Figure \ref{fig:thz1} that both excited tryptophan chromophores (Trp132 and Trp211) undergo clearly significant enhancements across the entire low-frequency THz range. This likely reflects subtle changes in the local environment of each tryptophan on the picosecond timescale consistent with previous studies\cite{singer2007}. The changes in the THz spectra associated with the autocorrelations of the two tryptophans upon photoexcitation are over a factor of 10 larger than what one would expect from random fluctuations. Based on the current simulation data, if relevant to the NEQ observation, these large absorption differences are prime candidates for energy funnelling to stimulate the BSA collective vibrational mode. Indeed, without a UV LED exciting these tryptophans in the NEQ experiment, no collective phononic mode is observed in the BSA solution. \cite{MPettiniprivcomm}

In addition, as shown in Figure \ref{fig:thz2}, there are statistically significant cross-correlation changes in the few-THz regime for the entire protein with ions and with water, and for both tryptophans with water. In recent years, there has been much interest in the relevance of low-frequency THz modes to ionic solutions \cite{funkner2012watching}. Our results suggest the possibility that photoexcitation or other local charge reorganization processes may induce global changes in the low-frequency THz modes of protein, ions, and water. 

Still, the physical relevance of the ``oscillations'' observed in the autocorrelation spectra---especially for Trp132-AC and Trp211-AC in Figure \ref{fig:thz1}---remains unclear at the moment. On the one hand, it may arise from the treatment of the linewidth analysis (focusing on the bare intensity differences at the frequency resolution applied), or on the other hand, it may be a signature of charge-phonon correlations that are beyond the scope of this paper. Further simulations and experiments are thus warranted to determine the sensitivity and importance of these thermodynamic averaging procedures in NEQ settings.

These $2\sigma$-significant values of different autocorrelations and cross-correlations from Figures \ref{fig:thz1} and \ref{fig:thz2} are summarized in Table S1. Near the experimentally observed reference frequency ($\nu \approx 0.3$ THz), besides autocorrelations from protein, water, and the whole system, cross-correlations between Trp132 and water, and between Trp211 and ions, appear integral to mediating the multiscale response of the whole system to photoexcitation.  All in all, the emerging picture from this analysis suggests that the few-THz signal involves a complex and multiscale response involving two photoexcited tryptophan amino acids, the entire BSA protein, ions, and water. Although the spectra are constructed from equilibrium simulations, the synopsis in Table S1 indicates the possibility of injecting energy into a wide variety of different modes involving water, ions, and protein, with the individual components and collective interactions supporting the maintenance of a coherent elastic vibration constrained by the protein bulk mechanical properties. Since the few-THz modes involve low-frequency collective vibrations, in the remainder of the manuscript we will quantify the molecular reverberations associated with changes in the protein, ions, and solvent that are triggered by UV photoexcitation of the two tryptophans in BSA.

\subsection{Collective structural reorganization of BSA protein}


The BSA protein simulated in this work has two tryptophans (Trp132 and Trp211), which are both substantially exposed to the solvent. The upper right panel of Figure \ref{fig:THzspec} shows a snapshot of the protein taken from our simulations, where the two tryptophans are highlighted with appropriate labels. 

\begin{figure}
    \centering
    \includegraphics[width=7cm]{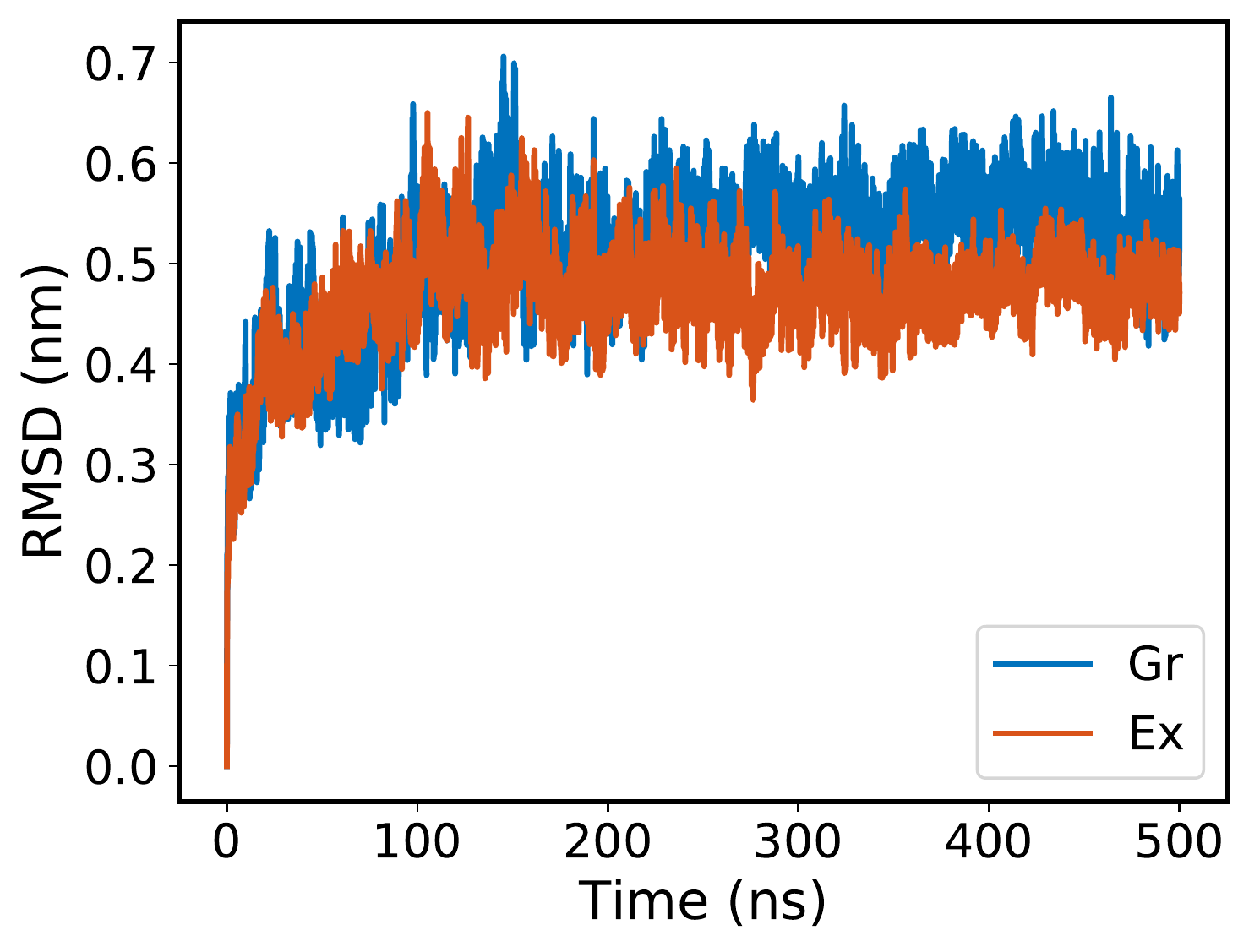} ~~
    \includegraphics[width=7cm]{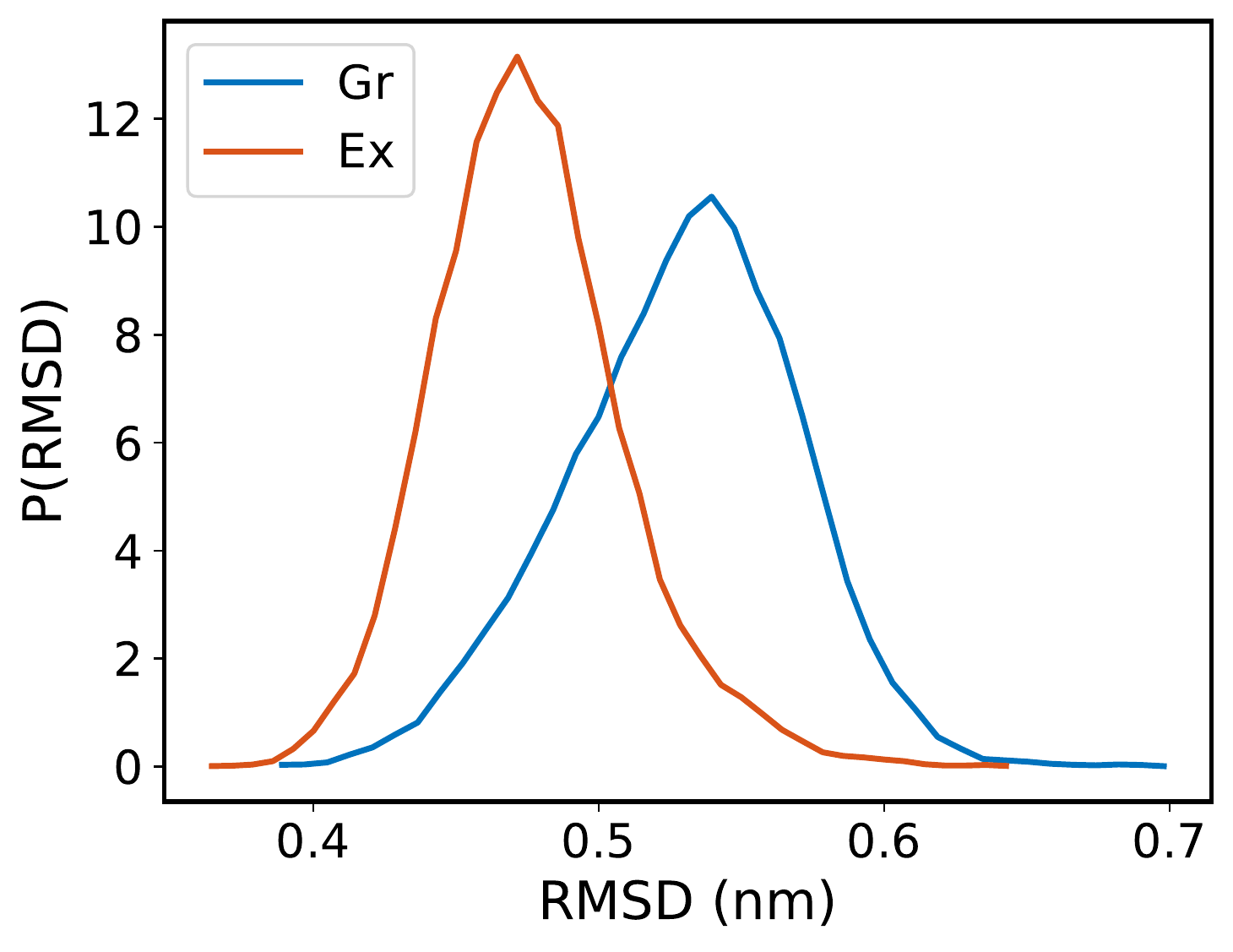} \\
    \includegraphics[width=7cm]{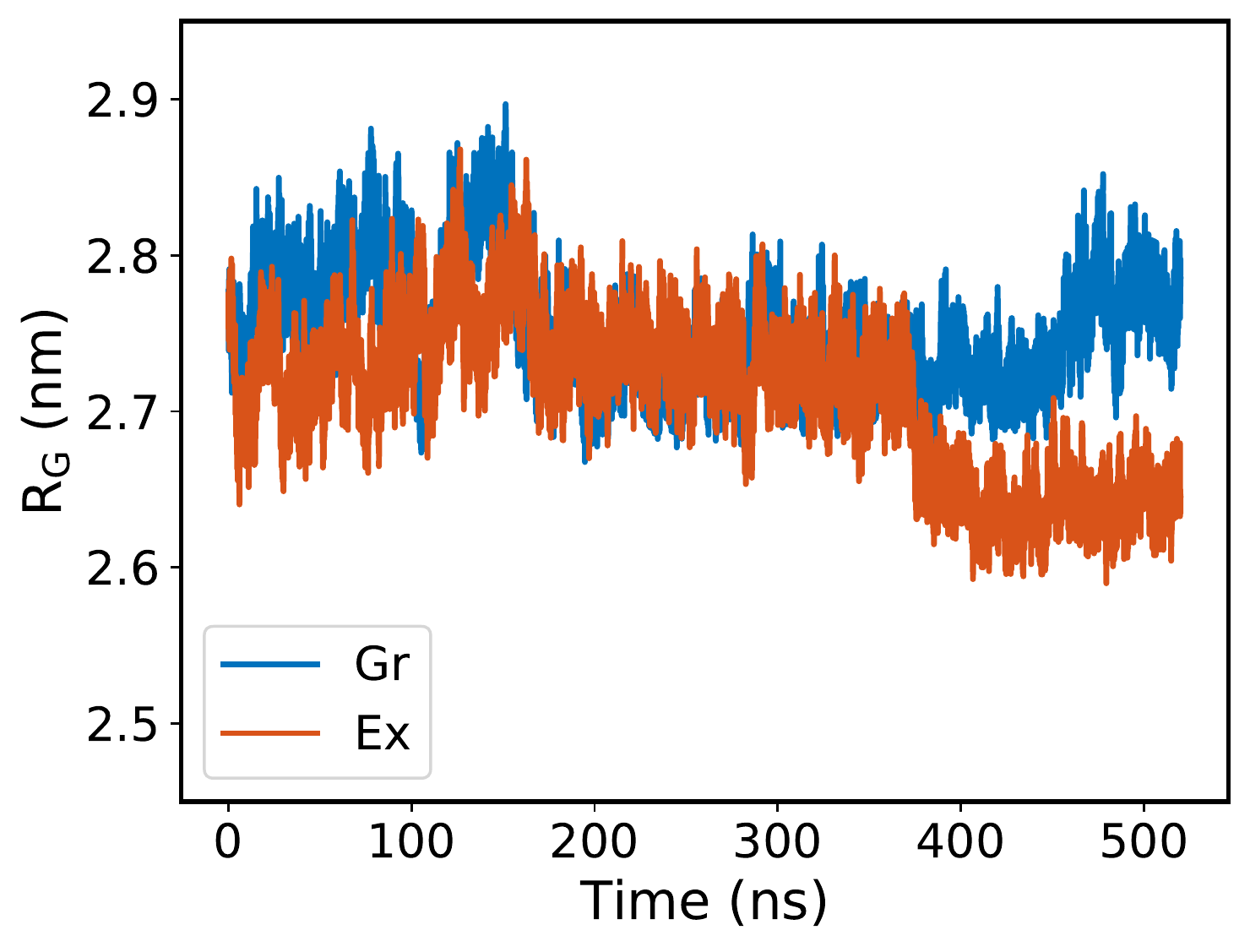} ~~
    \includegraphics[width=7cm]{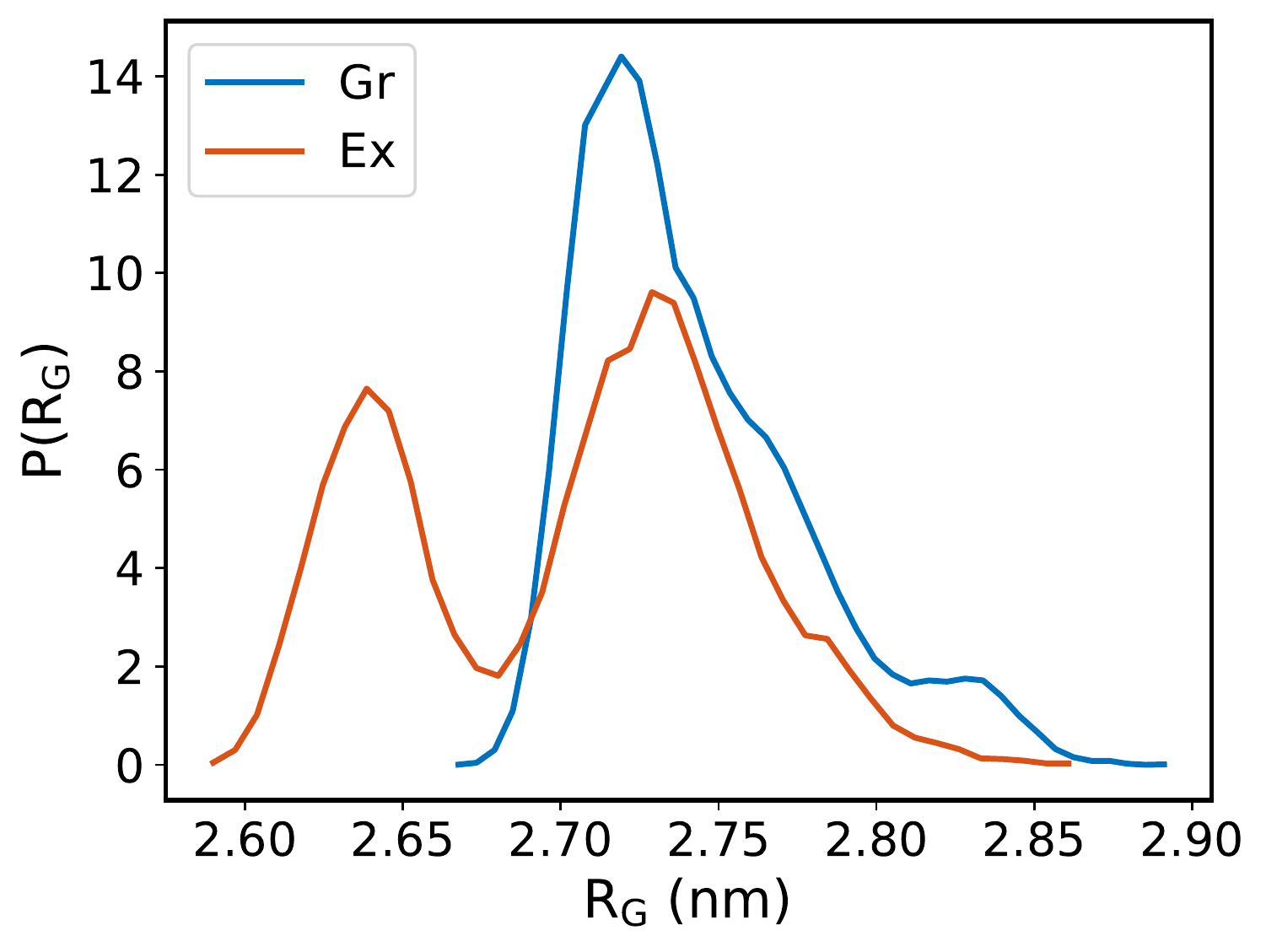} \\
    \includegraphics[width=7cm]{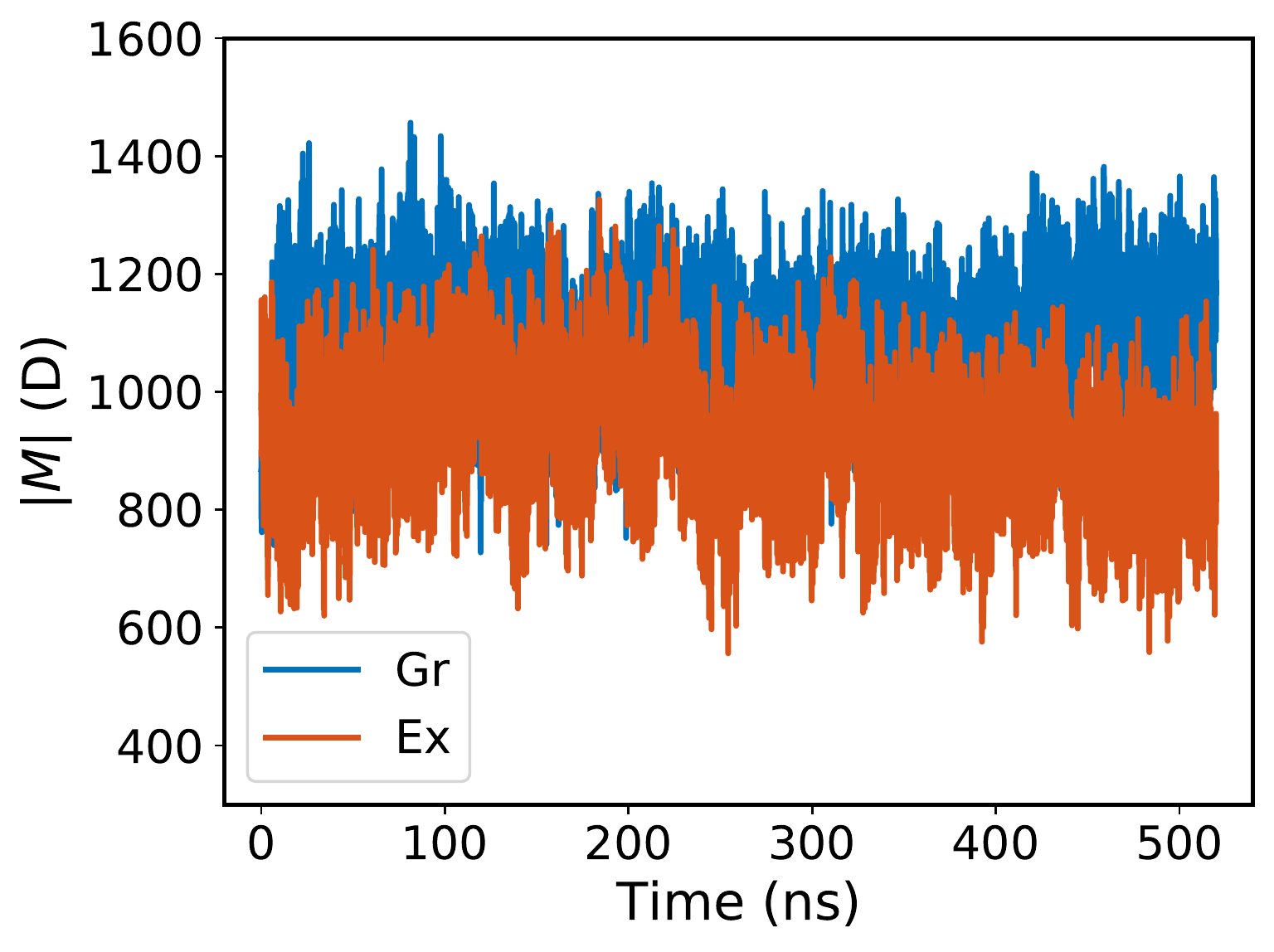} ~~
    \includegraphics[width=7cm]{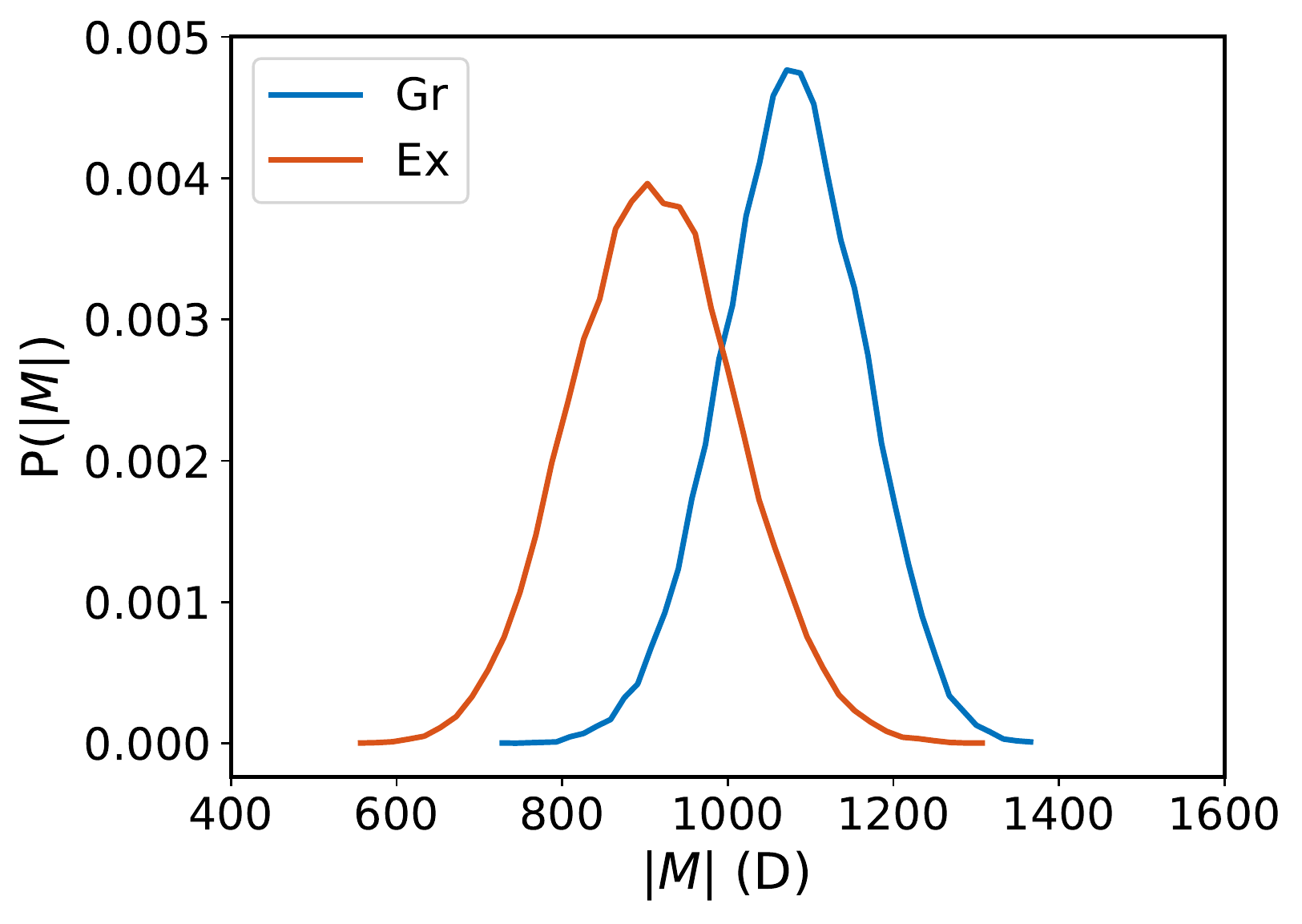}
    \caption{Time evolution and  distributions of different properties of BSA protein in aqueous solution: root-mean-square deviation (RMSD), radius of gyration ($R_G$), and the  total dipole moment ($|M|$), in the ground (Gr) and excited (Ex) state. The distributions are shown for equilibrated time series data after 100 ns. The equilibrium time is chosen based on the RMSD plot, which can also be confirmed by its distribution. The $R_G$ plot shows a conformational change in the excited state at about \SI{375}{ns}, which results in a bimodal behavior in the distribution and may be indicative of protein oscillations occurring on microsecond timescales. Total dipole moments of the protein---not including ionic contributions from the aqueous environment---are quite robust during the whole simulation time, with a lower mean value  but larger distribution width for the excited state due to the larger dipole fluctuations caused by photoexcitation.}
    \label{fig:rmsd_rg}
\end{figure}

In order to examine how the photoexcitation perturbs the mechanical behaviors of the protein, we determined the root-mean-square deviation (RMSD), radius of gyration ($R_G$), and the total dipole moment ($M$) of the protein, in both the ground and excited state. The left column of panels in Figure \ref{fig:rmsd_rg} shows the time series associated with these three quantities, while the right column shows the corresponding probability distributions. We observe that for both the ground- and excited-state trajectories it takes about 100 ns to equilibrate the system from the initial configuration. Thus in order to construct the histograms we do not include the first 100 ns.

Interestingly, we observe that the photoexcitation results in an overall rigidification of the protein as indicated by changes in the RMSD and $R_G$ distributions.  The time series of the $R_G$ shows that there are protein conformational changes occurring on a rather long time scale. In particular, we see a transition occurring at about 375 ns in the excited-state trajectory that results in a reduction of the $R_G$ by approximately 1 \AA. This change in the $R_G$ leads to the bimodal structure of the probability distribution in the excited state. Furthermore, the distribution of the RMSD in the excited state is slightly more narrow than that in the ground state. 

The bottom row of panels in Figure \ref{fig:rmsd_rg} shows the time series and corresponding distributions of the total dipole moment for the protein itself. Consistent with the findings of the RMSD and $R_G$, the UV photoexcitation of BSA  results in a significant reduction of the dipole moment of the protein (by about 20\%). It is interesting to note that the changes in $R_G$ are not manifested in $|M|$, which in the excited state retains a unimodal character. It is clear, however, that these types of atomistic details would not be captured by heuristic models and therefore are complemented by experiments \cite{lechelon_experimental_2022} 
examining the source of long-range dipole interactions as electrodynamic fluctuations of proteins dressed by their local ionic environment.

The changes in the RMSD, $R_G$, and $|M|$ upon photoexcitation suggest that there is a rather significant change in the structure of the protein. However, these quantities only probe the global structure of the protein. In order to provide a more refined view of the molecular reorganization and collective changes in BSA upon photoexcitation, we determined the difference in the so-called distance contact map between all pairs of amino acids in the protein. The contact map quantifies the local environment around each amino acid in terms of the neighbors in close proximity. Specifically, we sampled frames from the molecular dynamics time-series every 10 ps after equilibration for 420 ns (42000 frames) and determined the fraction of frames in which amino acids $i$ and $j$, for all amino acid pairs, are separated by less than 5 \AA. This is done for both the excited- and ground-state simulations, after which the difference between the two fractions is taken yielding the difference contact map. The choice of this cutoff length has been shown to provide a good balance of probing short-to-medium range correlations in protein structure in previous studies\cite{RivaltaE1428,eastpalermo2019,NegreE12201}. 

\begin{figure}
    \centering
     \includegraphics[scale=0.10]{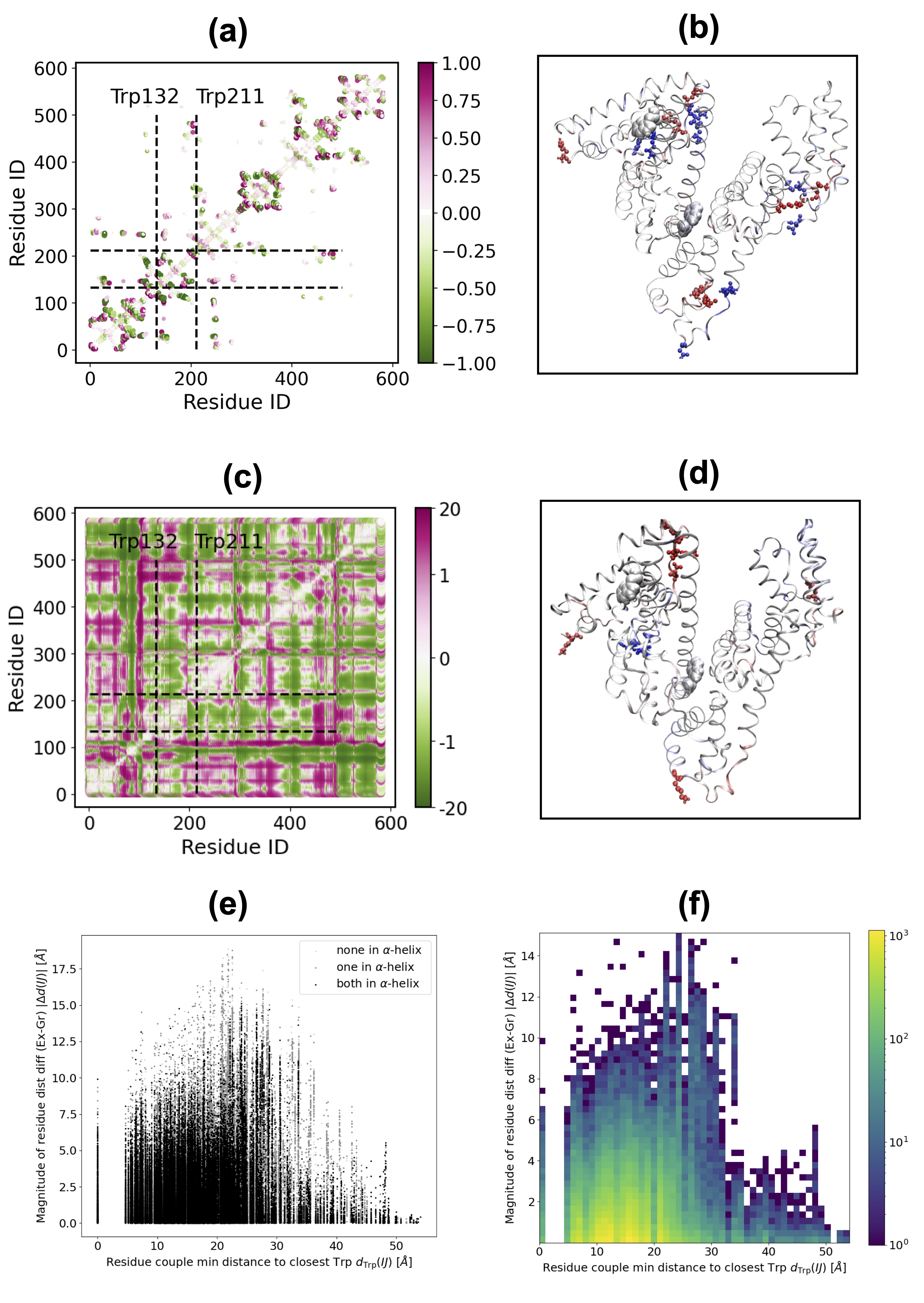}
    \caption{(a,b) The left panel shows the difference in the contact map between the excited- and ground-state simulations of the BSA aqueous solution. The presence of green and purple spots located across many residues, and not only along the vertical black lines (corresponding to the two tryptophans), indicates there are changes in the local geometries of amino acids throughout the protein. The right panel visually depicts this global dynamic restructuring.
    (c,d) 
    Variation of BSA average  conformations between excited and ground states, giving a complementary static picture to the dynamic behaviors shown in the upper panels.  
    (e,f) The left panel displays a correlogram between the minimum distance $d_{\rm Trp}(IJ)$ of a residue pair to the closest tryptophan and the magnitude of residue distance difference $| \Delta d(IJ)|$ between representative configurations for the excited and ground states of the BSA solution. The right panel displays a two-dimensional histogram
    representing in color scale the number of unique residue pairs with a given $|\Delta d(IJ)|$ and $d_{\rm Trp}(IJ)$, whose residues both belong to a BSA $\alpha$-helix. See main text for further details.}
    \label{fig:contactmap} 
\end{figure}

Figure \ref{fig:contactmap}(a) illustrates the difference contact map between excited- and ground-state simulations. As indicated above, the first 100 ns of equilibration were not included in this analysis. For clarity in the left panel, the dashed horizontal (vertical) lines colored in black indicate the amino acid indices of the two tryptophans. The presence of a green spot along one of these lines indicates those amino acids that move further away from the tryptophan on average over the sampled frames in the excited state (fewer contacts) relative to those in the ground state, while a purple spot corresponds to amino acids that move closer to the tryptophan on average over the sampled frames in the excited state (more contacts) relative to those in the ground state. It is worth noting that we observe changes in the rest of the protein as observed by the presence of green and purple spots across many amino acids in the system.

In order to quantify the net change in the total number of contacts, we sum over all the changes for each amino acid. To visualize these changes, in Figure \ref{fig:contactmap}(b), amino acids undergoing an increase in the number of contacts are colored in red while those that decrease in the number of contacts are shown in blue. For the purposes of visualization, we focus on only those contacts changing by more than 70\%. Interestingly, we observe that these regions are spread all over the protein. The distance of the tryptophans to these regions of the protein ranges from between one to over six nanometers and shows that there is a rather long-range structural reorganization upon photoexcitation.

The contact map analysis presented 
above provides insights into the
the difference in fluctuations of the BSA
configuration between photoexcited 
and ground states, but the contact map analysis 
introduces an arbitrary length scale to define 
the contact frequency. 
To obtain complementary information on the global 
BSA configuration induced by the tryptophan
photoexcitation, we examine the difference
$\Delta d(IJ) = d_{\rm Ex}(IJ) - d_{\rm GR}(IJ)$
of the 
residue distance matrices between two configurations
representative of the photoexcited and ground state. The distance $d_{Ex(Gr)}(IJ)$ between 
residues have been defined as the pairwise distance between the geometric barycenters of each residue in a configuration averaged over the
equilibrated $NVT$ time-series data presented in 
Figure \ref{fig:rmsd_rg}, after the $100$-ns equilibration for the excited (ground) state. 
Such a choice seems natural to encode in a 
single configuration, a representative whole of information on the 
thermodynamic equilibrium properties both in the ground and in 
the excited state, at the price of neglecting 
higher-order moments of the distribution and, as 
a result, losing information on dynamical
properties.

The results have been
reported in Figure \ref{fig:contactmap}(c), which affirms how the photoexcitation of
the tryptophans deeply affects the global structure of 
the protein, with a rearrangement of mutual residue distances
even for residue pairs not involving the tryptophans themselves. Moreover, it is 
interesting to notice that darkened regions, 
corresponding to the most important conformational changes, 
form a ``checkerboard" pattern parallel to the dashed 
Trp132 and Trp211 lines. Such darkened patterns
correspond to flexible and mobile secondary structure motifs (coils and turns), and their presence suggests the propagation of a structural rearrangement effect from the tryptophan centers to the rest of the protein.
Figure \ref{fig:contactmap}(d) shows a possible
encoding of the information contained in the 
difference matrix of Figure \ref{fig:contactmap}(d) into a residue-dependent scalar field $\phi(I)=\sum_{J} \Delta d(IJ)$ over the entire protein structure, analogous to Figure \ref{fig:contactmap}(b). 
The red spots show the residues for which 
$\phi(I)$ is large and positive: we observe how 
the photoexcitation of tryptophans induces a
configuration rearrangement of residue barycenters in sites both near and also far from the tryptophans.

In order to quantify this behavior more precisely,
a correlation plot has been reported in Figure
\ref{fig:contactmap}(e) between the 
the magnitude of the difference in mutual 
residue distances $|\Delta d(IJ)|$ and the 
distance of the residue in each pair closest to a 
tryptophan in the ground state defined as
\begin{equation}
d_{\rm Trp}(IJ)= \min_{K \in \{\rm Trp131, Trp211\}} (d_{\rm Gr}(IK),d_{\rm Gr}(JK)) \,.
\end{equation}
With such a definition, the residue pairs with 
$d_{\rm Trp}(IJ)=0$ are the ones containing at 
least one tryptophan. The residue pairs have been divided into three
groups: the pairs whose residues both belong to
an $\alpha$-helix (black points), 
the pairs where only one residue belongs to an
$\alpha$-helix (dark gray points), and the pairs whose
residues do not belong to any $\alpha$-helix (light gray points).
Such a classification has been introduced because the 
$\alpha$-helices are generally more rigid than other secondary-structure motifs \cite{tang1, tang2}, especially more so than the flexible coil and linker segments in BSA.
Still, the correlogram analysis reveals quite 
high values for the magnitude of residue distance 
difference ($|\Delta d(IJ)|>10 \,\SI{}{\angstrom}$) for 
some pairs belonging to $\alpha$-helices.
This can be interpreted as a manifestation of the 
effects of the photoexcitation on the BSA global
configuration, including more rigid 
structures such as the $\alpha$-helices. In fact, we can see from Figure \ref{fig:contactmap}(f) that
 a significant fraction ($\sim 4 \%$) of all pairs whose residues both belong to an $\alpha$-helix ($N=409$, out of 581 BSA residues, resulting in $N(N-1)/2=83436$ unique $\alpha$-helical residue pairs) and having $d_{\rm Trp}(IJ)\leq \SI{20}{\angstrom}$ show a magnitude of residue distance difference $|\Delta d(IJ)|>4 \,\SI{}{\angstrom}$ (the smallest length of an amino acid \cite{ching1989evaluation}) due to the photoexcitation of the tryptophans.

Numerous previous studies have shown, using similar models, that the optical excitation of tryptophan can induce a wide variety of protein structural changes, such as the movement of positively charged lysine groups due to the change in the strength of pi-cation interactions\cite{hassanalikwk2006} as well as perturbations to the protein secondary structure\cite{singer2007}. The preceding analysis in Figure \ref{fig:contactmap} provides a simpler way to pinpoint these structural perturbations. Moreover, these collective structural changes are likely to be important for tuning the THz modes discussed earlier.

In order to understand better how the fluctuations of each amino acid in the protein respond to optical excitations, we determined the difference in the root-mean-square fluctuations (RMSF) between the ground- and excited-state simulations (see Figure S2). Interestingly, the response of the protein is characterized by some amino acids incurring an enhancement in their fluctuations, while others become more restricted. The right panel in Figure S2 provides a visual depiction of the changes in RMSF. We see clearly that, contrary to some expectations, the amino acids undergoing large changes in RMSF do not necessarily involve those that are proximal to the optically excited chromophores.

\subsection{Photoinduced reorganization of ions and water}

Besides the collective structural changes in the protein, the terahertz spectra also suggest that there are important contributions coming from the reorganization of both ions and water molecules. Several previous studies have shown with time-dependent fluorescence Stokes shift experiments that the optical response involves protein, ionic, and aqueous contributions\cite{hassanali2016,hassanali2017,callis1994,fursecorcelli2008}. In addition, there have been several experimental measurements showing the importance of low-frequency ThZ modes that are sensitive to fluctuations of ion related modes\cite{havenith2009ion,havenith2012,havenith2019}. 


In Figure S3 we show the pair correlation functions between the two tryptophan residues and the sodium/chloride ions for both ground and excited states. The left panel shows the distribution functions for Trp132, and the right panel is for Trp 211. For Trp 132, there appear to be subtle changes occurring in the ion density upon photoexcitation, which occur over several nanometers around the excited chromophore. In the case of Trp211, the changes appear to be much more drastic, especially for the chloride ions within one nanometer. Visual examination of the environment of Trp211 shows that there is a cluster of positively charged arginine residues in close proximity, which attract more anions. Upon photoexcitation, the structural reorganization that occurs throughout the protein reduces the accessibility of these positively charged residues to the chloride ions.

The perturbations in the ion densities around the two tryptophans prompted us to examine whether there are any changes in the solvent exposure of the tryptophans. Toward this end, we determined the solvent-accessible surface area (SASA) of all the amino acids following photoexcitation. Specifically, the distribution of the change in the SASA is shown in the left panel of Figure S4. While it is peaked around zero, there are amino acids whose SASA increases or decreases upon photoexcitation, particularly those at the solvent boundary. A visual illustration of the change in the SASA is shown in the right panel of Figure S4, which shows that there are regions of the protein close to and far away from the two excited tryptophans that undergo changes in their solvent exposure.

It is important to note that if ion distributions change significantly near the protein, as suggested in Figure S3, there could be substantial contributions to the effective dipole moment arising from this recruitment. Indeed, notwithstanding the probability distributions for $|M|$ in Figure \ref{fig:rmsd_rg} that only consider protein contributions without surrounding ions, further investigations are warranted to verify if recruitment of ions around the rigidified protein in the excited state could be the source of a ``giant dipole" enhancement upon photoexcitation. 

\subsection{Photoinduced dynamic correlations via eigenvector centrality analysis}

In the last decade, there has been significant development in the confluence of ideas from both network and information theory in order to understand the nature of fluctuations in biological systems\cite{NegreE12201,Lange2006,eastpalermo2019,Nierzwicki2021,RivaltaE1428,Amaro2012}. Within the context of protein dynamics, the eigenvector centrality provides a manner to quantify how dynamically correlated an amino acid is with other components of its networked environment. Batista and co-workers recently employed the centrality measure to examine the dynamical changes that occur during effector-enzyme binding\cite{NegreE12201}.  Details of how the eigenvector centrality is computed can be found in the Methods section.

\begin{figure}
    \centering
    \includegraphics[scale=0.6]{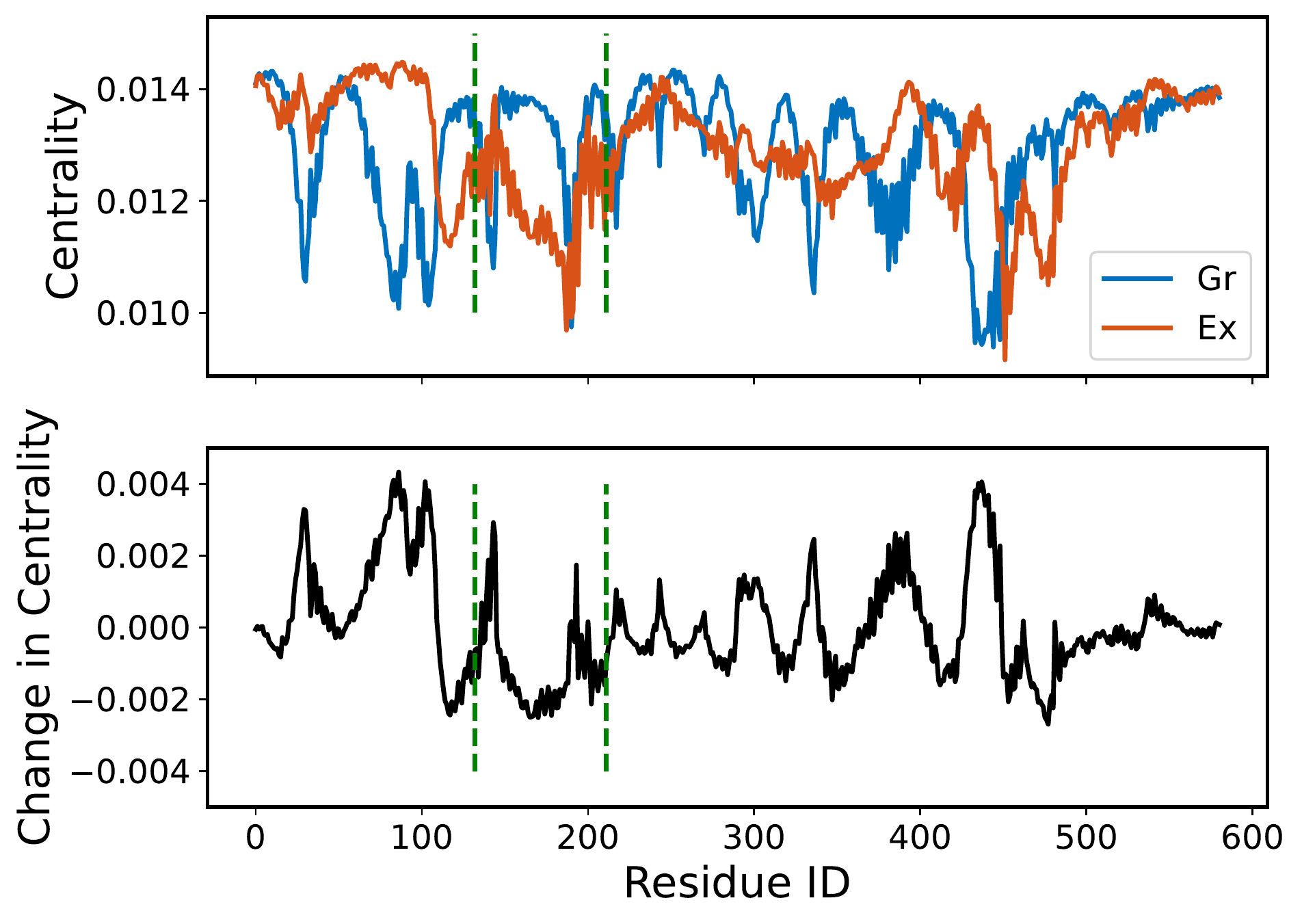} \\ \vspace{0.6cm}
    \includegraphics[scale=0.5]{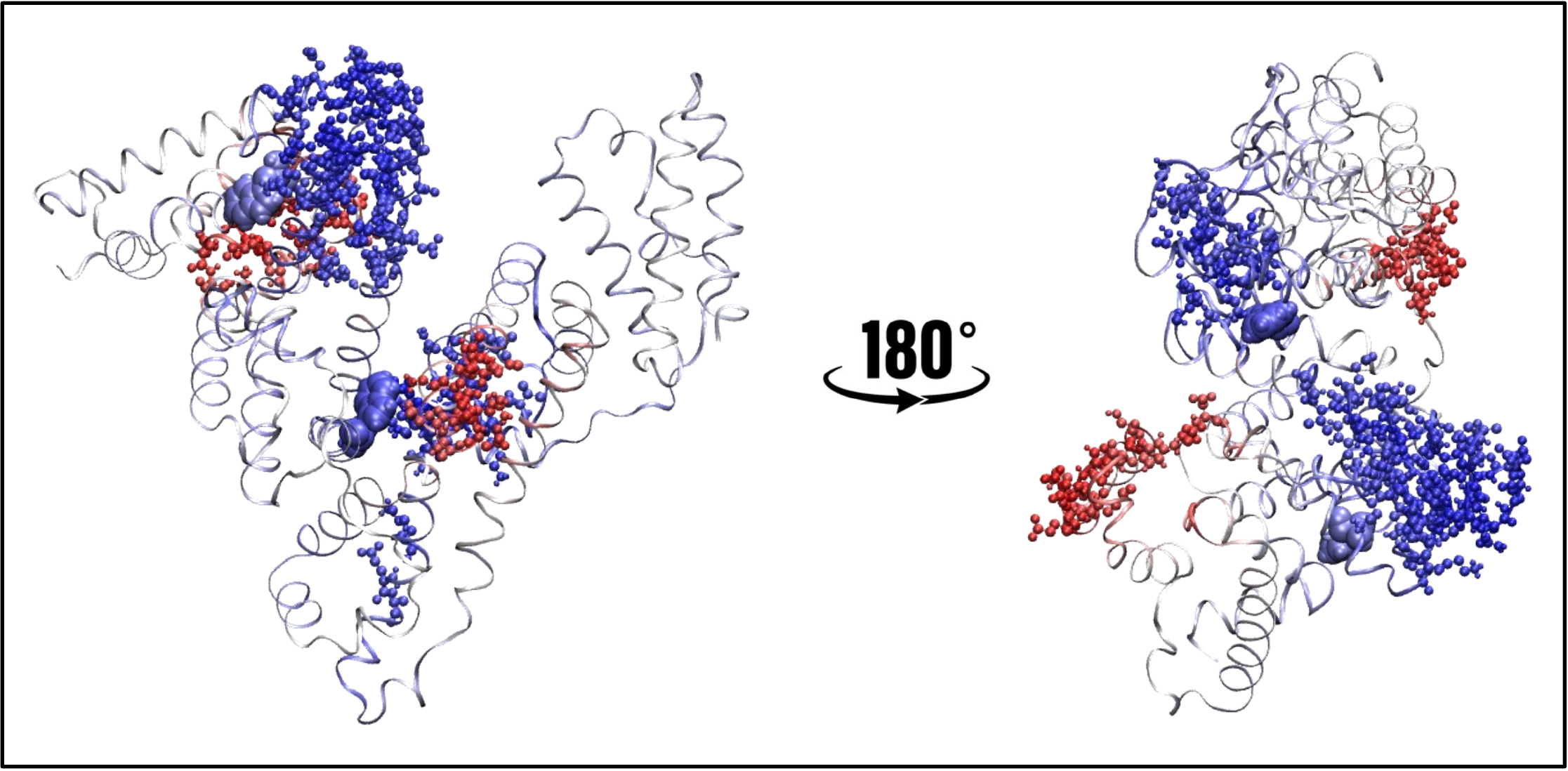}
    \caption{Upper panel: Eigenvector centrality distribution for BSA in the ground state (Gr) and photoexcited state (Ex) (blue and orange lines, respectively). Middle panel: Centrality differences (Ex - Gr)  as a function of the residue index. Lower panel:  Centrality differences (Ex - Gr) plotted over the 3D protein structure. Red and blue residues indicate the regions with the largest (above 70\% of the values for other residues) gain and loss, respectively, of centrality upon photoexcitation.}
    \label{fig:centrality}
\end{figure}

The upper panel of Figure \ref{fig:centrality} presents a visual depiction of the change in the centrality of BSA upon tryptophan photoexcitation. Specifically, we employ the centrality defined in Equation \ref{eq:11}, where the E$_c$ vector has a unit norm. Areas colored in red correspond to regions of the protein that become more correlated with the major collective fluctuation modes of BSA, while those in blue are regions that lose correlation with the principal fluctuation modes of the protein. Interestingly, and in agreement with the structural behavior, the dynamical changes highlighted by this analysis are not limited to the proximity of the excited tryptophans but rather delocalized, indicating that the global dynamics of BSA is modified upon tryptophan photoexcitation. Figure \ref{fig:centrality} shows that the main photoexcitation-induced dynamical perturbation of BSA occurs at the nexus of its three globular domains. This suggests that tryptophan photoexcitation will strongly impact the inter-domain vibrational modes of the protein, and the changes in centrality around this nexus would be highly correlated to spheroidal breathing motions of the whole protein that have been predicted \cite{bastrukov1994} and observed \cite{nardecchia2018out, cheatum2020} at $\sim 0.3$ THz. Further analysis tracking which residues specifically contribute to such vibrational modes---projecting dynamical correlations into that basis---would be the subject of future work. 


\section{Conclusions and Discussion}

Recent THz spectroscopy experiments on photoexcited proteins in aqueous solution \cite{nardecchia2018out, lechelon_experimental_2022} have allowed observation of a phenomenology suggesting the activation of giant electric dipole oscillations. In particular, it has been observed that the THz absorption features of an ionic solution of BSA protein marked with Alexa488 fluorophore changes when the sample is excited simultaneously by a \SI{488}{\nano\meter} laser and a UV LED centered at \SI{255}{\nano\meter}. In these experiments, the appearance of a distinct peak in the absorption spectrum at \SI{314}{\giga\hertz} has been interpreted as the activation of a phononic collective breathing mode in BSA. In fact, a theoretical model predicts that the lowest-frequency breathing mode for an elastic sphere having the same mechanical properties of BSA (i.e., shear modulus and bulk density) is at $\SI{308}{\giga\hertz}$. Such optomechanical excitations would be responsible for the proposed giant dipole oscillations that could play an important role in intermolecular signaling at the cellular scale and beyond. For instance, long-range electrodynamic interactions \cite{lechelon_experimental_2022} may be established among resonant oscillators carrying a large electric dipole, contributing to the efficient recruitment of cognate partners to a local vicinity in space and time for biochemical reactions to occur. 

The activation of such giant electric dipole oscillations in BSA is due to a two-step process. The first step is the downconversion of energy through vibrational relaxation in photoexcited fluorophores, which thereby act as an energy source for the mechanical degrees of freedom in the protein. The second step is the concentration of phononic excitations in the lowest-frequency vibrational mode according to a nonlinear mechanism analogous to Fr\"{o}hlich condensation in quantum or classical out-of-thermal equilibrium systems. Progress in understanding the first step has been made with hybrid quantum mechanical MD simulations showcasing the coupling of tryptophan excited states (L$_a$) to charge transfer (CT) states in proteins \cite{callis2011}, which increases the likelihood of nonradiative decay into vibrational channels when the CT states overlap in energy with the excited states. Our work makes a significant contribution toward dissecting the diversity of modes that could be involved in the second step.

While the microscopic details of the activation of these giant dipole oscillations still remain poorly understood, our THz and structural/dynamical analyses presented in this paper have illuminated some of the main features and ``levers'' of the multiscale optomechanical transduction process in an atomistic model of solvated BSA. We have shown that in the critical sub-THz regime where the nonequilibrium observation at $\sim 0.3$ THz was made, a variety of statistically significant modes arise dynamically from auto- and cross-correlations of interacting system components (tryptophan residues, entire protein, ions, water, whole system), upon optical excitation of the two BSA tryptophan residues during equilibrium MD simulation. The autocorrelation frequency of the entire protein, and the cross-correlations for Trp211-ions and for Trp132-water, are the closest in frequency to the experimental value, suggesting that a truly collective and strongly coupled interaction between protein molecules and their ionic solvent produces the observed difference in THz absorption between excited and ground states.  

To explain the structural reverberations emerging from the optical excitation, we have shown the global restructuring of residue contacts throughout the protein, even more than six nanometers away from the photoexcited tryptophans. Commensurately, we observe stark changes in the radius of gyration of the BSA protein, which exhibits a pronounced conformational change in the excited state near the end of the $0.5$-$\mu$s simulation, suggesting that protein oscillations might manifest on longer timescales. We have also demonstrated that significant reorganizations take place in the ions and water around the tryptophan residues and at the protein-solvent boundary, by tracking radial distribution functions for sodium and chloride ions and solvent-accessible surface areas for all the amino acids. Such rearrangements could have a considerable impact on the total effective dipole moment of the protein in ionic solvent. Taken together, these data suggest drastic deviations of our simulated system from the standard expectations of linear response theory. 

This work is intended to be the first step toward investigating the origins of the experimentally observed spectroscopic THz features using atomistic simulations. Atomistic simulations of electromagnetic radiation at different frequencies interacting with biological materials, in our case a solution of BSA protein, present many challenges, including realization of the precise binding sites and orientations for the Alexa488 fluorophores, the details of vibrational relaxation, and downconversion of energy from electronically excited Alexa488 and tryptophan to the protein vibrational and solvent degrees of freedom. Here we specifically focused on the problem of how photoexcitation of the two tryptophans in BSA protein can result in a collective response of the environment. The aim is twofold: From one side we have investigated how photoexcitation of tryptophans of native BSA (not marked with Alexa488) in ionic solution affects the mechanical and electronic response properties of the protein and its environment, and on the other side we have studied how local changes in the biomolecule could have global effects on the atomistic dynamics, for instance in the generation of new collective normal modes and the centrality evolution of components in the network.

Despite this progress in our understanding of the emergence of collective optomechanical vibrational phenomena, many aspects remain to be clarified in open systems pumped far from thermal equilibrium. Our current molecular dynamics simulations of photoexcited BSA solution are done at equilibrium. Indeed, there have been several experimental and theoretical studies that have examined the role of nonequilibrium (NEQ) effects within the context of deviations from linear response theory. The nonlinear response of so-called ``soft modes'' (i.e., a subset of polar low-frequency phonons) to a resonant external driving field and the related local-field corrections have remained largely unexplored, especially in proteins. In aspirin the correlation of sidechain rotovibrational modes in the few-THz frequency range with collective oscillations of $\pi$ electrons drives the system into the nonperturbative regime of light-matter interaction, even for moderate strengths of the driving field \cite{aspirin2017}. This nonlinear response---and its role in light-driven, mechanical-structural phase transitions---originates from the strong coupling of electronic and vibrational degrees of freedom. This effect has also been described for Fr\"ohlich polarons \cite{ultrafastnonlinear2007}, where the impulsive movement of an electron in the highly nonlinear regime induces coherent phonons that persist for several hundred femtoseconds. These technical issues may reveal fundamental concerns at the center of debates about the breakdown of ergodicity and linear response in biological systems.

Such nonlinearities in protein vibrational dynamics have also been well described by Davydov and others in the theory of solitonic propagation along $\alpha$-helices \cite{davydov2010}. This theory poses significant challenges for MD simulations of nonlinear dissipative systems \cite{prigogine2017} that are ubiquitous in biology, and which reflect the flow of energy under NEQ conditions. Indeed, the overdamping of sidechain rotovibrational modes in fully equilibrated MD simulations is related to the suppression of transient states that may mask the emergence of pronounced conformational changes occurring on much longer timescales. Our results suggest that these protein conformational changes in MD simulations can occur, in spite of overdamping, after several hundred nanoseconds (see Figure \ref{fig:rmsd_rg}, radius of gyration), and that residues in $\alpha$-helices are significantly affected (see Figure \ref{fig:contactmap}, bottom correlograms). These changes would likely take on greater prominence should additional nonlinear effects be included. Short of this, we interpolate in this work between strictly equilibrium simulations and NEQ conditions by the inclusion of electronically excited tryptophan residues in our thermodynamically equilibrated BSA simulations. In this way, our results motivate future NEQ simulations with nonlinear inclusions.

The breakdown of linear response theory may be closely related to the deviations observed from ergodic behavior in systems of biological relevance. UV fluorescence upconversion experiments combined with NEQ MD simulations have observed deviations from the linear response approximation for the relaxation dynamics of photoexcited tryptophan in water \cite{chergui2010}. It has also been demonstrated that metabolic activities drive the biological milieu toward non-ergodicity far from thermodynamic equilibrium, resulting in increased cytoplasmic fluidization that allows larger components to escape their local environment anomalously and explore larger regions of the cellular compartment \cite{jacobs-wagner2014}. The limitations of equilibrium MD simulations allude to the need for interfacing quantum excited states in biomolecules with an atomistic bath that includes a nonlinear dissipative term describing the flow of energy between the two. Indeed, pioneering work \cite{klinman1999,klinman2011,klinman2019} has demonstrated the role of such thermal conduits in kinetic activation of catalytically efficient states in enzyme dynamics. These perspectives and our results will motivate future, more realistic NEQ MD simulations and additional experiments to characterize the extent, limits of control, and applications of Fr\"{o}hlich-like phenomena and optomechanical transduction in biology.

\begin{acknowledgement}
This work has been supported in part by the Defense Advanced Research Projects Agency. This research used resources of the Argonne Leadership Computing Facility, which is a DOE Office of Science User Facility supported under contract DE-AC02-06CH11357. This research also used resources of the Oak Ridge Leadership Computing Facility, which is a DOE Office of Science User Facility supported under Contract DE-AC05-00OR22725. Finally, we would like to acknowledge conversations with Matthias Heyden and Marco Pettini in advancing this work. 
\end{acknowledgement}

\section{Competing Interest Information}
All authors declare that there are no competing financial or other interests.

\section{Data Sharing Plans}
Input files to run the GROMACS simulations, trajectories generated for computation of terahertz spectra, and the codes/scripts used for analysis of the resulting data can be obtained from the authors upon reasonable request. After publication, these materials will be provided on the GitHub repository. 


\clearpage
\providecommand{\latin}[1]{#1}
\makeatletter
\providecommand{\doi}
  {\begingroup\let\do\@makeother\dospecials
  \catcode`\{=1 \catcode`\}=2 \doi@aux}
\providecommand{\doi@aux}[1]{\endgroup\texttt{#1}}
\makeatother
\providecommand*\mcitethebibliography{\thebibliography}
\csname @ifundefined\endcsname{endmcitethebibliography}
  {\let\endmcitethebibliography\endthebibliography}{}


\clearpage
\section
  {Supporting Information} 
  

\renewcommand{\thefigure}{S\arabic{figure}}
\setcounter{figure}{0} 

\renewcommand{\thetable}{S\arabic{table}}
\setcounter{table}{0} 

\begin{figure}[h!]
    \centering
    \includegraphics[scale=0.6]{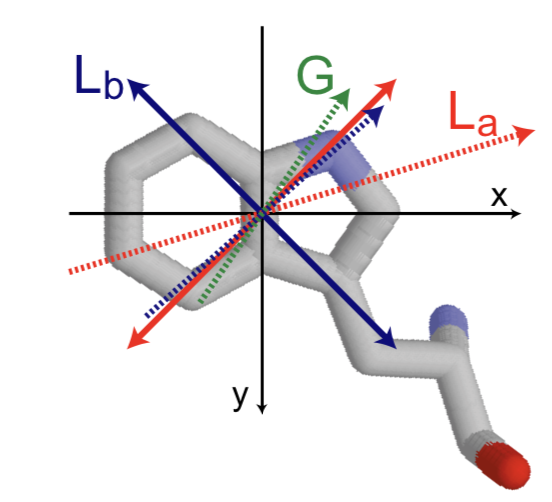}
    \caption{Dipole moments of tryptophan. Full arrows represent the transition dipole moments, and dashed ones represent the permanent dipole moments. The ground state G is depicted in green. Dipole moments of the excited state $\text{L}_\text{a}$ ($\text{L}_\text{b}$) are depicted in red (blue). Reproduced from S. Schenkl et al., \textit{Science} \textbf{309,} 917–920 (2005), supplementary information.}
    \label{fig:LaLb}
\end{figure}

\begin{figure}
    \centering
    \includegraphics[scale=0.55]{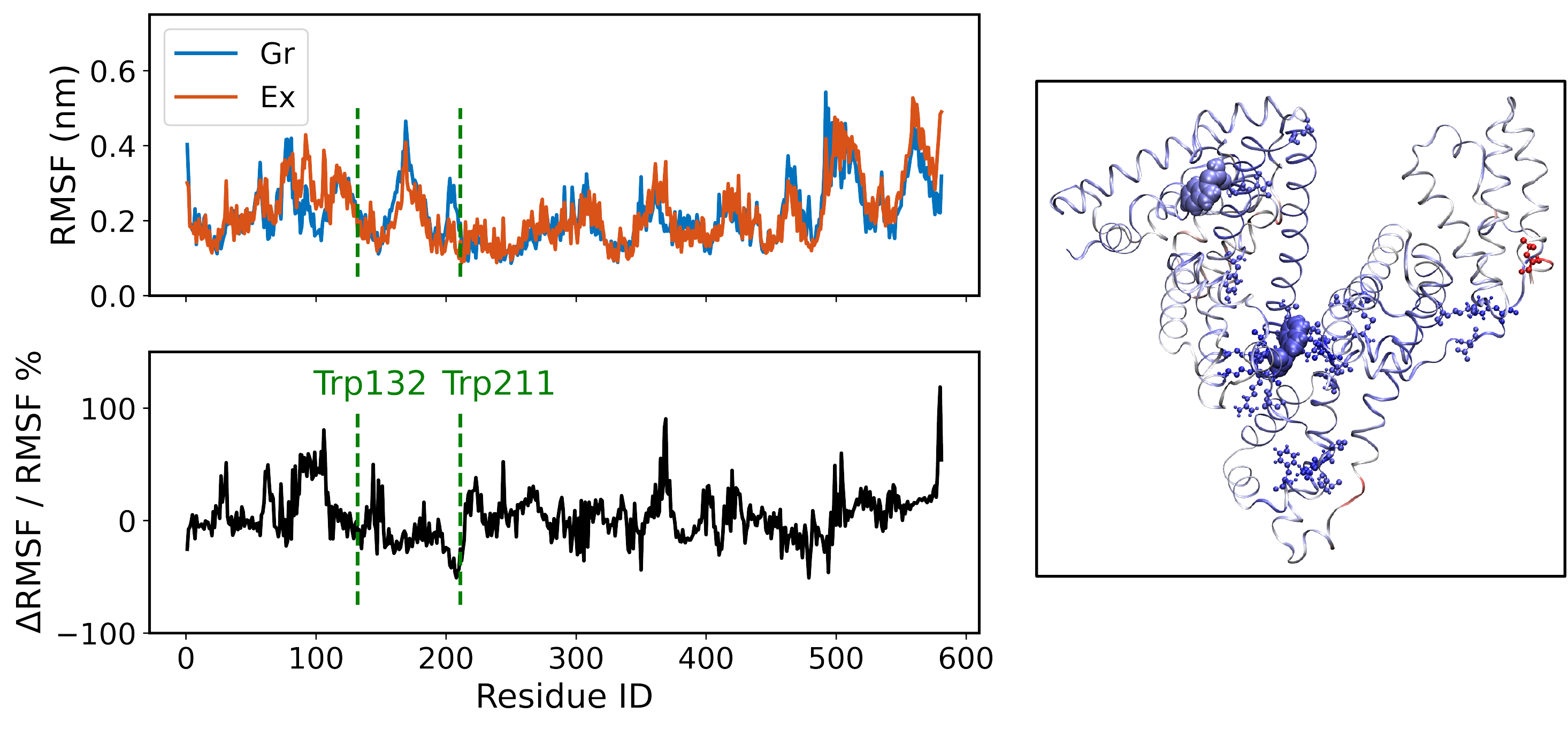}
    \caption{Root mean square fluctuations (RMSF) and percent change for BSA upon photoexcitation and after thermodynamic equilibrium. The right configuration shows the color-coding of the change in RMSF shown on top of the BSA structure. The residues with 70\% increase (decrease) in their values are depicted in red (blue) colors. The two tryptophans are enlarged in size for clarity. The large number of blue residues spanning the protein highlights that the protein becomes more rigid due to the photoexcitation.}
    \label{fig:rmsf_change}
\end{figure}

\begin{figure}
    \centering
    \includegraphics[scale=0.385]{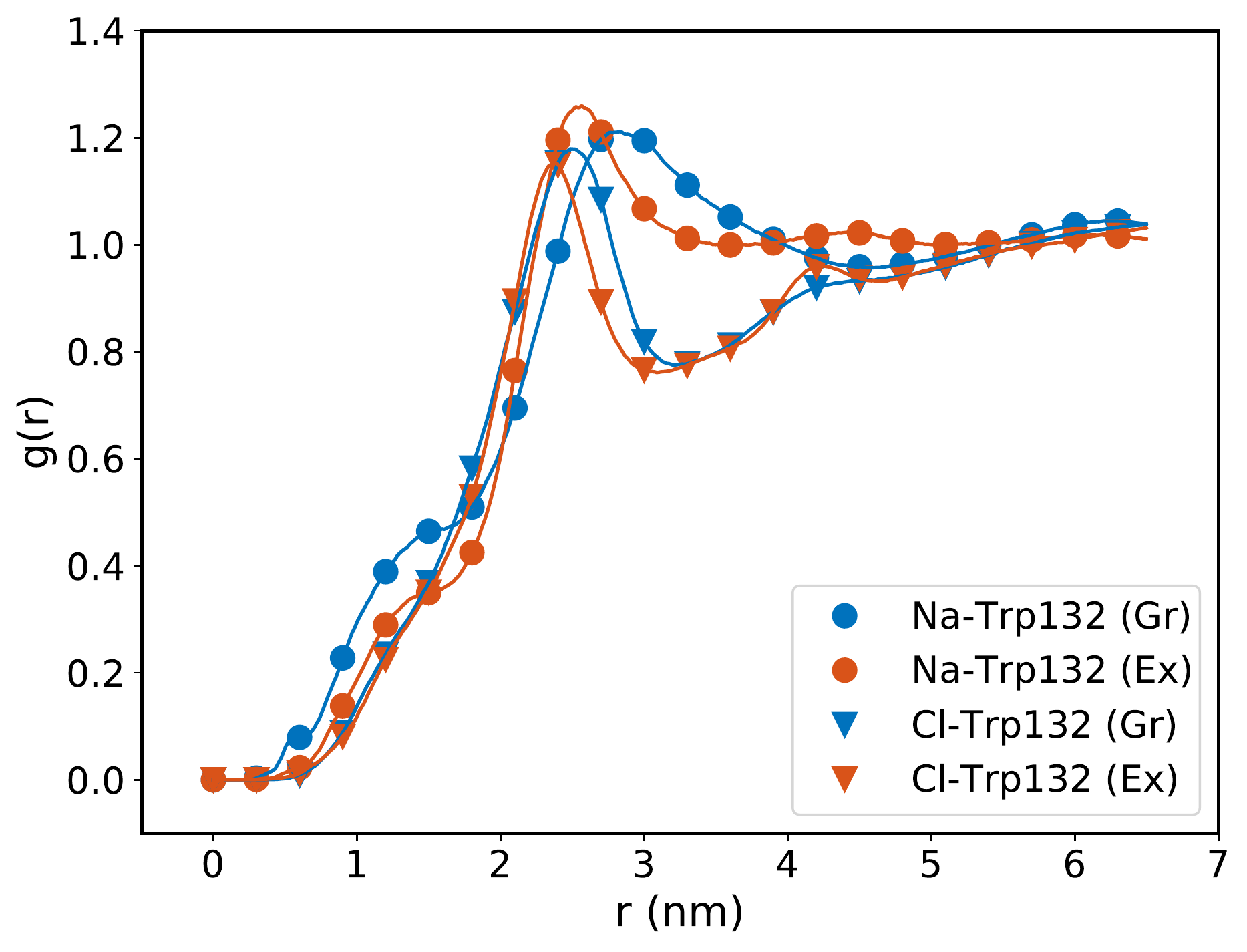} 
    \includegraphics[scale=0.385]{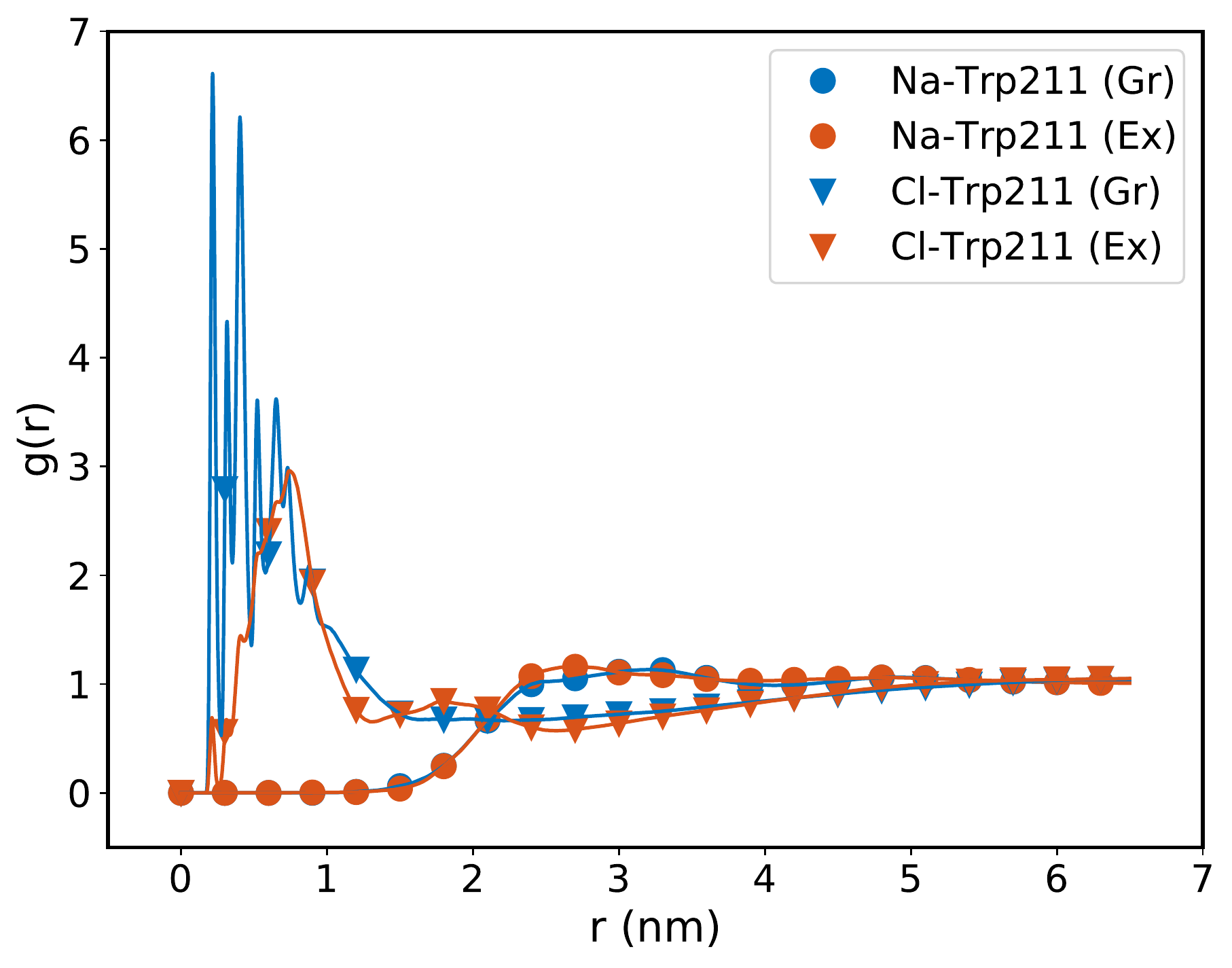}
    \caption{Radial distribution function of the sodium (Na, circle) and chloride (Cl, triangle) ions around the two tryptophans of BSA in the ground (Gr, blue) and excited (Ex, orange) state. The proximity of the cations and anions for the two tryptophans is apparently inverted. The sodium ions are generally equally correlated to either tryptophan, while the chloride ions on the other hand are clearly more correlated to Trp211. Furthermore, photoexcitation results in subtle changes in the ion densities around Trp132, while in the case of Trp211 the chloride ion distributions undergo more pronounced changes upon photoexcitation which, as mentioned in the main text, likely arise from conformational changes reducing accessibility of a proximal cluster of positively charged arginine residues.}
    \label{fig:rdf}
\end{figure}

\begin{figure}
    \centering
    \includegraphics[scale=0.6]{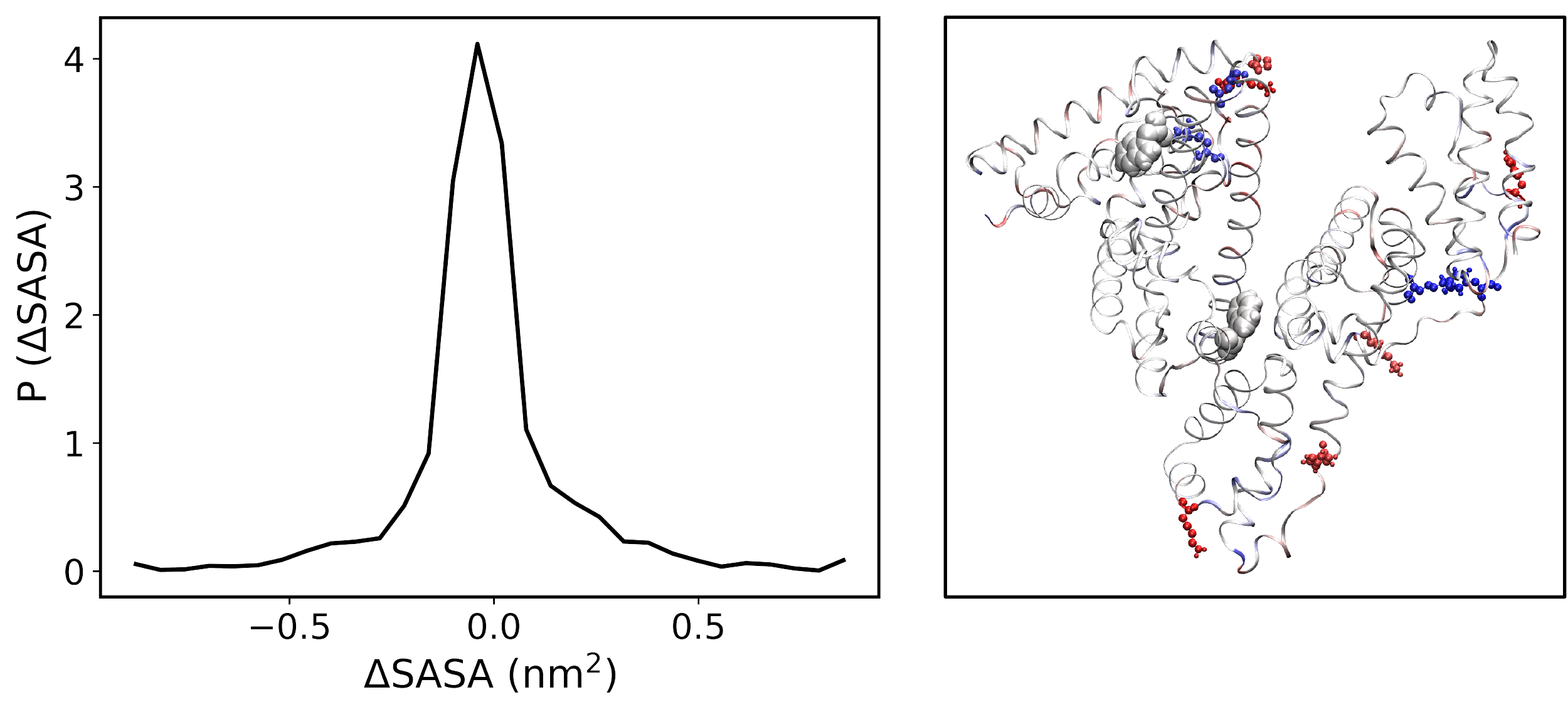}
    \caption{ Probability distribution of change in the solvent accessible surface area (SASA) of BSA upon photoexcitation.
    The right panel illustrates the results mapped onto the BSA protein by red (blue) color showing a 70\% increase (decrease) in the SASA due to photoexcitation. The change in SASA is seen for the boundary residues, while the residues not at the surface do not change their behavior as dramatically.}
    \label{fig:sasa_change_hist}
\end{figure}

\begin{table}[h!]
\begin{tabular}{|l|c|c|c|c|}
\hline
Channel & $\nu [\mathrm{THz}]$ & $|\Delta \alpha (\nu)|/\sigma(\nu)$   & $\Delta \alpha(\nu)$ [a.u.] & Conf. of $|\Delta \alpha (\nu)|$ \\
\hline
Protein-AC & $0^{*}$ & 2.9 & 0.0614 & 98.6 \% \\
\hline
System-AC & \bf 0.125 & 2.1 & 8.96 & 96.4 \%\\
\hline
Water-AC & \bf 0.125 & 2.2 & 9.20 & 97.2 \%\\
\hline
Trp132-CC-Water & \bf 0.125 & 2.2 & -0.0299 & 97.2 \% \\
\hline
Protein-AC & \bf 0.250 & 2.9 & 0.0907 & 99.6 \% \\
\hline
Trp211-CC-Ions & \bf 0.375 & 2.1 & 0.00349 & 96.4 \% \\
\hline
Trp132-CC-Water & \bf 0.375 & 2.9 & -0.0396 & 99.6 \% \\
\hline
Trp211-CC-Ions & 0.626 & 2.5 & -0.0381 & 98.8 \% \\
\hline
Trp132-CC-Water & 0.626 & 2.5 & -0.0381 & 98.8 \% \\
\hline
Trp132-CC-Water & 1.13 & 2.3 & -0.0432 & 97.9 \% \\
\hline
Protein-AC & 1.38 & 2.0 & 0.198 & 95.9 \% \\
\hline
 Ions-AC & 1.38 & 2.7 & 0.584 & 99.3 \% \\
 \hline
Protein-CC-Ions & 1.38 & 2.2 & -0.474 & 97.2 \% \\
\hline
Protein-AC & 1.50 & 2.5 & 0.274 & 98.8 \% \\
\hline
Trp132-CC-Water & 1.63 & 2.1 & -0.0413 & 96.4 \% \\
\hline
Trp211-CC-Water & 1.75 & 2.2 & 0.0680 & 97.2 \% \\
\hline
Trp211-CC-Water & 1.88 & 3.6 & -0.119  & 99.97  \% \\
\hline
Protein-CC-Water & 2.25 & 2.4 & -2.82 & 98.4 \% \\
\hline
\end{tabular}
\caption{
Synopsis of the autocorrelation (AC) and cross-correlation (CC) data points exceeding $2\sigma$-statistical significance for the different channels reported in Figures 2 and 3 in the main text, highlighting the subset of points (in bold) nearest to the experimentally observed reference frequency at $\nu \approx 0.3$ THz. The * for the zero-frequency value highlights that the average has been taken over a finite interval $T$ in the time domain, so the amplitude of the spectrum effectively includes information on frequencies smaller than $1/T$ in principle. 
For each data point the frequency $\nu$, the corresponding value of $|\alpha(\nu)|$ in units of $\sigma(\nu)$ defined in Equation 3 in the main text, the difference of the absorption coefficients between the photoexcited and ground states $\Delta \alpha(\nu)=\alpha_{\text{Ex}}(\nu)-\alpha_{\text {Gr}}(\nu)$, and the statistical confidence interval associated with $|\Delta \alpha(\nu)|$ are listed.}
\end{table}

\end{document}